\documentclass[journal]{IEEEtran}

\usepackage[numbers,sort&compress]{natbib}
\usepackage{url}
\usepackage{bm}
\usepackage{amsfonts}
\usepackage{amsmath}
\usepackage{amssymb}
\usepackage{amsthm}
\usepackage{color}

\usepackage{array,color}
\usepackage{tabularx}
\usepackage{multirow}
\usepackage{booktabs}
\usepackage{makecell}  

\usepackage{graphicx}  
\usepackage{subfig}
\usepackage{epstopdf}
\usepackage{graphics}
\usepackage{epsfig}
\usepackage{psfrag}
\usepackage{overpic}

 \usepackage{stfloats}
 \usepackage{float}   

 \usepackage{indentfirst} 

\usepackage{gensymb}

\begin{document}

\newcommand {\Data} [1]{\mbox{${#1}$}}  

\newcommand {\DataN} [2]{\Data{\Power{{#1}}{{{#2}}}}}  
\newcommand {\DataIJ} [3]{\Data{\Power{#1}{{{#2}\!\times{}\!{#3}}}}}  

\newcommand {\DatassI} [2]{\!\Data{\Index{#1}{\!\Data 1},\!\Index{#1}{\!\Data 2},\!\cdots,\!\Index{#1}{\!{#2}}}}  
\newcommand {\DatasI} [2]{\Data{\Index{#1}{\Data 1},\Index{#1}{\Data 2},\cdots,\Index{#1}{#2},\cdots}}   
\newcommand {\DatasII} [3]{\Data{\Index{#1}{{\Index{#2}{\Data 1}}},\Index{#1}{{\Index{#2}{\Data 2}}},\cdots,\Index{#1}{{\Index{#2}{#3}}},\cdots}}  

\newcommand {\DatasNTt}[3]{\Data{\Index{#1}{{#2}{\Data 1}},\Index{#1}{{#2}{\Data 2}},\cdots,\Index{#1}{{#2}{#3}}} } 
\newcommand {\DatasNTn}[3]{\Data{\Index{#1}{{\Data 1}{#3}},\Index{#1}{{\Data 2}{#3}},\cdots,\Index{#1}{{#2}{#3}}} } 

\newcommand {\Vector} [1]{\Data {\mathbf {#1}}}
\newcommand {\Rdata} [1]{\Data {\hat {#1}}}
\newcommand {\Tdata} [1]{\Data {\tilde {#1}}} 
\newcommand {\Udata} [1]{\Data {\overline {#1}}} 
\newcommand {\Fdata} [1]{\Data {\mathbb {#1}}} 
\newcommand {\Prod} [2]{\Data {\prod_{\SI {#1}}^{\SI {#2}}}}  
\newcommand {\Sum} [2]{\Data {\sum_{\SI {#1}}^{\SI {#2}}}}   
\newcommand {\Belong} [2]{\Data{ {#1} \in{}{#2}}}  

\newcommand {\Abs} [1]{\Data{ \lvert {#1} \rvert}}  
\newcommand {\Mul} [2]{\Data{ {#1} \times {#2}}}  
\newcommand {\Muls} [2]{\Data{ {#1} \! \times \!{#2}}}  
\newcommand {\Mulsd} [2]{\Data{ {#1} \! \cdot \!{#2}}}  
\newcommand {\Div} [2]{\Data{ \frac{#1}{#2}}}  
\newcommand {\Trend} [2]{\Data{ {#1}\rightarrow{#2}}}  
\newcommand {\Sqrt} [1]{\Data {\sqrt {#1}}} 
\newcommand {\Sqrtn} [2]{\Data {\sqrt[2]{#1}}} 

\newcommand {\Power} [2]{\Data{ {#1}^{\TI {#2}}}}  
\newcommand {\Index} [2]{\Data{ {#1}_{\TI {#2}}}}  

\newcommand {\Equ} [2]{\Data{ {#1} = {#2}}}  
\newcommand {\Equs} [2]{\Data{ {#1}\! =\! {#2}}}  
\newcommand {\Equss} [3]{\Equs {#1}{\Equs {#2}{#3}}}  

\newcommand {\Equu} [2]{\Data{ {#1} \equiv {#2}}}  

\newcommand {\LE}[0] {\leqslant}
\newcommand {\GE}[0] {\geqslant}
\newcommand {\NE}[0] {\neq}
\newcommand {\INF}[0] {\infty}
\newcommand {\MIN}[0] {\min}
\newcommand {\MAX}[0] {\max}

\newcommand {\Funcfx} [2]{\Data{ {#1}({#2})}}  
\newcommand {\Funcfzx} [3]{\Data{ {\Index {#1}{#2}}({#3})}}  
\newcommand {\Funcfnzx} [4]{\Data{ {\Index {\Power{#1}{#2}}{#3}}({#4})}}  
\newcommand {\SI}[1] {\small{#1}}
\newcommand {\TI}[1] {\tiny {#1}}
\newcommand {\Text}[1] {\text {#1}}

\newcommand {\VtS}[0]{\Index {t}{\Text {s}}}
\newcommand {\Vti}[0]{\Index {t}{i}}
\newcommand {\Vt}[0]{\Data {t}}
\newcommand {\VLES}[1]{\Index {\tau} {\SI{\Index {}{ \Text{#1}}}}}
\newcommand {\VLESmin}[1]{\Index {\tau} {\SI{\Index {}{ \Text{min\_\Text{#1}}}}}}
\newcommand {\VAT}[0]{\Index {\Vector A}{\Text{Time}}}
\newcommand {\VPbus}[1]{\Index {P}{\Text{Bus-}{#1}}}
\newcommand {\VPbusmax}[1]{\Index {P}{\Text{max\_Bus-}{#1}}}

\newcommand {\EtS}[2]{\Equs {\Index {t}{\Text {s}}}{#1} {#2}}
\newcommand {\Eti}[2]{\Equs {\Index {t}{i}}{#1} {#2}}
\newcommand {\Et}[2]{\Equs {t}{#1} {#2}}
\newcommand {\EMSR}[2]{\Equs {\Index {\tau} {\SI{\Index {}{ \Text{MSR}}}}}{#1} {#2}}
\newcommand {\EMSRmin}[2]{\Equs {\Index {\tau} {\SI{\Index {}{ \Text{MSR}}}}}{#1} {#2}}

\newcommand {\EAT}[2]{\Equs {\Index {\Vector A}{\Text{Time}}}{#1} {#2}}
\newcommand {\EPbus}[3]{\Equs {\Index {P}{\Text{Bus-}{#1}}}{#2} \Text{ #3}}
\newcommand {\EPbusmax}[3]{\Equs {\Index {P}{\Text{max\_Bus-}{#1}}}{#2} \Text{ #3}}

\newcommand {\Vgam}[1]{\Index {\gamma}{#1}}
\newcommand {\Egam}[2]{\Equs {\Vgam{#1}}{#2}}

\newcommand {\Emu}[2]{\Equs {{\mu}{#1}}{#2}}
\newcommand {\Esigg}[2]{\Equs {{\sigma}^2{#1}}{#2}}

\newcommand {\Vlambda}[1]{\Index {\lambda}{#1}}

\newcommand {\VV}[1]{\Index {\Vector V}{#1}}
\newcommand {\Vv}[1]{\Index {\Vector v}{#1}}
\newcommand {\Vsv}[1]{\Index {v}{#1}}
\newcommand {\VX}[1]{\Index {\Vector X}{#1}}
\newcommand {\VsX}[1]{\Index {X}{\SI{\Index {}{#1}}}}
\newcommand {\Vx}[1]{\Index {\Vector x}{\SI{\Index {}{#1}}}}
\newcommand {\Vsx}[1]{\Index {x}{\SI{\Index {}{#1}}}}
\newcommand {\VZ}[1]{\Index {\Vector Z}{#1}}
\newcommand {\Vz}[1]{\Index {\Vector z}{\SI{\Index {}{#1}}}}
\newcommand {\Vsz}[1]{\Index {z}{\SI{\Index {}{#1}}}}
\newcommand {\VIndex}[2]{\Index {\Vector {#1}}{#2}}
\newcommand {\VY}[1]{\Index {\Vector Y}{#1}}
\newcommand {\Vy}[1]{\Index {\Vector y}{#1}}
\newcommand {\Vsy}[1]{\Index {y}{\SI{\Index {}{#1}}}}

\newcommand {\VRV}[1]{\Index {\Rdata {\Vector V}}{#1}}
\newcommand {\VRsV}[1]{\Index {\Rdata {V}}{#1}}
\newcommand {\VRX}[1]{\Index {\Rdata {\Vector X}}{#1}}
\newcommand {\VRx}[1]{\Index {\Rdata {\Vector x}}{\SI{\Index {}{#1}}}}
\newcommand {\VRsx}[1]{\Index {\Rdata {x}}{\SI{\Index {}{#1}}}}
\newcommand {\VRZ}[1]{\Index {\Rdata {\Vector Z}}{#1}}
\newcommand {\VRz}[1]{\Index {\Rdata {\Vector z}}{\SI{\Index {}{#1}}}}
\newcommand {\VRsz}[1]{\Index {\Rdata {z}}{\SI{\Index {}{#1}}}}
\newcommand {\VTX}[1]{\Index {\Tdata {\Vector X}}{#1}}
\newcommand {\VTx}[1]{\Index {\Tdata {\Vector x}}{\SI{\Index {}{#1}}}}
\newcommand {\VTsx}[1]{\Index {\Tdata {x}}{\SI{\Index {}{#1}}}}
\newcommand {\VTsX}[1]{\Index {\Tdata {X}}{\SI{\Index {}{#1}}}}
\newcommand {\VTZ}[1]{\Index {\Tdata {\Vector Z}}{#1}}
\newcommand {\VTz}[1]{\Index {\Tdata {\Vector z}}{\SI{\Index {}{#1}}}}
\newcommand {\VTsz}[1]{\Index {\Tdata {z}}{\SI{\Index {}{#1}}}}
\newcommand {\VOG}[1]{\Vector{\Omega}{#1}}

\newcommand {\Sigg}[1]{\Data {{\sigma}^2({#1})}}
\newcommand {\Sig}[1]{\Data {{\sigma}({#1})}}

\newcommand {\Mu}[1]{\Data{{\mu} ({#1})}}
\newcommand {\Eig}[1]{\Data {\lambda}({\Vector {#1}}) }
\newcommand {\Her}[1]{\Power {#1}{\!H}}
\newcommand {\Tra}[1]{\Power {#1}{\!T}}

\newcommand {\VF}[3] {\DataIJ {\Fdata {#1}}{#2}{#3}}
\newcommand {\VRr}[2] {\DataN {\Fdata {#1}}{#2}}

\newcommand {\Tcol}[2] {\multicolumn{1}{#1}{#2} }
\newcommand {\Tcols}[3] {\multicolumn{#1}{#2}{#3} }
\newcommand {\Cur}[2] {\mbox {\Data {#1}-\Data {#2}}}

\newcommand {\VDelta}[1] {\Data {\Delta\!{#1}}}

\newcommand {\STE}[1] {\Fdata {E}{\Data{({#1})}}}
\newcommand {\STD}[1] {\Fdata {D}{\Data{({#1})}}}

\newcommand {\TestF}[1] {\Data {\varphi(#1)}}
\newcommand {\ROMAN}[1] {\uppercase\expandafter{\romannumeral#1}}
\newcommand {\Matrx}[3]{\mbox {\Data{\mathbf{#1}_{{#2}\!\times\!{#3}}}}}
\newcommand {\MatrxS}[2]{\mbox {\Data{\mathbf{#1}_{{#2}}}}}

\def \FuncC #1#2{
\begin{equation}
{#2}
\label {#1}
\end{equation}
}

\def \FuncCC #1#2#3#4#5#6{
\begin{equation}
#2=
\begin{cases}
    #3 & #4 \\
    #5 & #6
\end{cases}
\label{#1}
\end{equation}
}

\def \Figff #1#2#3#4#5#6#7{   
\begin{figure}[#7]
\centering
\subfloat[#2]{
\label{#1a}
\includegraphics[width=0.23\textwidth]{#4}
}
\subfloat[#3]{
\label{#1b}
\includegraphics[width=0.23\textwidth]{#5}
}
\caption{\small #6}
\label{#1}
\end{figure}
}

\def \Figffb #1#2#3#4#5#6#7#8#9{   
\begin{figure}[#9]
\centering
\subfloat[#2]{
\label{#1a}
\includegraphics[width=0.23\textwidth]{#5}
}
\subfloat[#3]{
\label{#1b}
\includegraphics[width=0.23\textwidth]{#6}
}

\subfloat[{#4}]{
\label{#1c}
\includegraphics[width=0.48\textwidth]{#7}
}
\caption{\small #8}
\label{#1}
\end{figure}
}

\def \Figffp #1#2#3#4#5#6#7{   
\begin{figure*}[#7]
\centering
\subfloat[#2]{
\label{#1a}
\begin{minipage}[t]{0.24\textwidth}
\centering
\includegraphics[width=1\textwidth]{#4}
\end{minipage}
}
\subfloat[#3]{
\label{#1b}
\begin{minipage}[t]{0.24\textwidth}
\centering
\includegraphics[width=1\textwidth]{#5}
\end{minipage}
}
\caption{\small #6}
\label{#1}
\end{figure*}
}

\def \Figf #1#2#3#4{   
\begin{figure}[#4]
\centering
\includegraphics[width=0.48\textwidth]{#2}

\caption{\small #3}
\label{#1}
\end{figure}
}

\definecolor{Orange}{RGB}{249,106,027}
\definecolor{sOrange}{RGB}{251,166,118}
\definecolor{ssOrange}{RGB}{254,213,190}

\definecolor{Blue}{RGB}{008,161,217}
\definecolor{sBlue}{RGB}{090,206,249}
\definecolor{ssBlue}{RGB}{200,239,253}

\title{Designing for Situation Awareness of Future Power Grids:  An Indicator System Based on Linear Eigenvalue Statistics of Large Random Matrices}

\author{Xing~He,  Robert~C. Qiu,~\IEEEmembership{Fellow,~IEEE}, Qian~Ai, ~\IEEEmembership{Member,~IEEE}, Lei~Chu, Xinyi~Xu, Zenan~Ling
\thanks{This work was partly supported by National Natural Science Foundation of China (No. 51577115 and No. 61571296).}
}

\maketitle

\begin{abstract}

Future power grids are fundamentally different from current ones, both in size and in complexity; this trend  imposes challenges for situation awareness (SA) based on classical indicators, which are usually model-based and deterministic.
As an alternative, this paper proposes a statistical indicator system based on \textit{linear eigenvalue statistics} (LESs)  of large random matrices:
1) from a data modeling viewpoint,  we build, starting from power flows equations, the random matrix models (RMMs) only using the real-time data flow in a statistical manner;
2) for a data analysis that is fully driven from RMMs, we put forward the high-dimensional indicators, called LESs that have some unique statistical features such as Gaussian properties;
and 3) we develop a three-dimensional (3D) power-map to visualize the system, respectively, from a high-dimensional viewpoint and a low-dimensional one.
Therefore, a statistical methodology of SA is employed; it conducts SA with a \textit{model-free and data-driven} procedure, requiring no knowledge of system topologies, units operation/control models, causal relationship, etc. This methodology has numerous advantages, such as sensitivity, universality, speed, and flexibility. In particular, its robustness against \textit{bad data}  is highlighted, with potential advantages in cyber security. The theory of big data based stability for on-line operations may prove feasible along with this line of work, although this critical development will be reported elsewhere.

\end{abstract}


\IEEEpeerreviewmaketitle

\section{Introduction}
\label{Intro}
\IEEEPARstart{T}{he} modern power gird is one of the most complex engineering systems in existence; the North American power grid is recognized as the supreme engineering achievement in the 20th century \cite{doe2003grid}. The complexity of the future's electrical grid is ever increasing: 1) the evolution of the grid network, especially the expansion in size;  2) the penetration of renewable/distributed resources, flexible/controllable electronic components, or even prosumers with dual load-generator behavior~\cite{grijalva2011prosumer}; and 3) the revolution of the operation mechanism, e.g., demand-side management. Also, the financial, the environmental and the regulatory constrains are pushing the electrical grid towards its stability limit.

All these driving forces demand a new prominence to the term \emph{situation awareness} (SA). The SA is essential for power grid security; inadequate SA is identified as one of the root causes  for the largest blackout in history---the 14 August 2003 Blackout in the United States and Canada \cite{us2004final}.

\subsection{Main Objective and Related Researches}
SA is a big topic; its time requirements are varying---from a few milliseconds for transient procedure, several minutes for long-time emergency of the line thermal ratings~\cite{wong1999ieee}, to months or even years for grid planning. This paper is mainly concerned about short-time phenomena and timely SA with real-time data flow.

There are numerous studies about utilizing phasor measurement units (PMUs) to improve wide-area monitoring, protection and control~\cite {phadke2008wide,terzija2011wide,xie2014dimensionality}. Data management for PMU-centric datasets is highly demanding and critical to the future grid. The advantage of efficient, PMU-based metrics is model-free and in real-time, without knowledge of network parameters or topology. Based solely on PMUs data, we argue that such an approach is well suitable for near-real-time applications. Singular Value Decomposition (SVD) is used in such large systems~\cite{lim2016svd}.
Giri~\cite{giri2012situation} points out that good advanced analytics and a visualization framework  is able to present the operator with a comprehensive, holistic portrayal of current grid condition.
Xu~\cite {xu2013power} initiates power disturbance data analytics to explore potential applications of power quality monitoring data; the mathematical foundations and working architectures are not included in his work.
Lim~\cite{lim2016svd}  proposes a voltage stability and conditioning monitor that is model-free and in real-time.
Simpson~\cite{simpson2016voltage} derives a closed-form condition under which a power network is safe from voltage collapse; the condition combines the complex structure of the network with the reactive power demands of loads to produce a node-by-node \emph{measure} of grid stress, a \emph{prediction} of the largest nodal voltage deviation, and an \emph{estimate} of the distance to collapse.

\subsection{Contribution and Previous Work }
\label{ContriNS}
Random Matrix Theory (RMT) tools are really underestimated in power girds---there are about 1,000 IEEE journals/magazines related to RMT from 2012 to 2015; most of those are in the field of information, communication, signal processing. RMT can also lead to lots of fruitful results in power systems---we have already proposed the first systematic architecture of RMT approach in power systems \cite{he2015arch}, and also given some practical applications related to the anomaly detection \cite{he2015arch} and the correlation analysis \cite{he2015corr}. These studies introduce the RMT to power grids and show some unique advantages; for instance, anomaly detection  using high-dimensional indicators (e.g., MSR\footnote{ Mean Spectral Radius, proposed in \cite{he2015arch}} $\tau{}_{\text{MSR}} : \tau_{\text{MSR}}= \tau_0$), is much more sensitive and effective than using classical one (e.g. voltage magnitude on node $i$ at time $j$: $0.94<v_{i,j}<1.06$). They lead to a conclusion that  the data-driven methodology, rather than the model-based alternative, is more suitable for a \textit{complex, large-scale interconnected grid} with massive data.

The previous work used a relative rough statistical model and a certain indicator, i.e.  $\tau_{\text{MSR}}$;
this paper continues with the line and gives a more concrete treatment---making full use of the massive data in order to obtain an indicator system for the timely SA. The core function of this indicator system is to tell  \emph{signals} from \emph{noises}:
\begin{itemize}
\item \emph{Noises}: conventional incidents, e.g.,  sample errors, and irregular power fluctuations of small loads/generators.
\item \emph{Signals}:  anomaly events, e.g., faults, network reconfigurations, and dramatic power changes on some bus.
\end{itemize}

  Although  Ring Law and Marchenko-Pastur (M-P) Law have already been proposed in previous two studies,   they are simply reviewed here, for the current paper to be self-contained. Some contributions are pointed out as follows:
1) Universality principle is needed to justify the use of RMT in engineering systems; 2) Linear Eigenvalue Statistics (LESs)---their definitions, statistical features, and engineering performance metrics against bad data\footnote{E.g., data errors, or even data losses in the core area}, as well as their comparison with classical indicators---are investigated as the major contribution; 3) Random matrix models (RMMs) and the hypothesis testing, as the pre-requisite for conducting data-driven SA via the LES indicator system, are introduced to tie together the physical system measurements with data analysis tools/methods.

\section{Background and Data-Driven Modeling for Power Grids}
\label{section: backgroundSA}

The nomenclature is given as Table \ref{tab:Nomenclature}.

\begin{table}[htbp]
\caption{Some Frequently Used Notations in the Theory}
\label{tab:Nomenclature}
\centering
\begin{tabularx}{0.48\textwidth} { l !{\color{black}\vrule width1pt} p{6.8cm} } 

\toprule[1.5pt]
\hline
\textit{Notations} & \textit{Meanings}\\

\Xhline{1pt}

\VX{},\Vx{},\Vsx,\Vsx{i\!,j} & a matrix, a vector, a single value, an entry of a matrix \\
\VRX{},\VRx{},\VRsx{} & hat: raw data\\
\VTX{},\VTx{},\VTsx{},\VTZ{} & tilde: intermediate variables, formed by normalization\\
\Data{N,T,c} & the numbers of rows and columns; \Equs{c}{N/T}\\
\Matrx {X}{N}{T}, \MatrxS {X}{N} & matrix $\mathbf{X}$ of size $N\! \times\! T$, $N \!\times\! N$\\

$\mathbf R$ & standard Gaussian random matrix, with i.i.d. entries $r_{i,j}$\\
$r$ & Gaussian random variable with mean 0 and variance 1\\

\Vector S & covariance matrix of \Matrx {X}{N}{T}:  $\mathbf{S}_N=1/T\mathbf{X}\mathbf{X}^H$  \\

%
%
%

$\mathbb {C},  \mathbb {R}$&  complex space, real space\\
%
%
%

%
\VLES{}{}, \VLES{MSR}{} &  linear eigenvalue statistics, mean spectral radius\\
$\varphi,\widehat{\varphi}$&the test function and its Fourier transformation\\
\Vlambda{} & the eigenvalue\\

\VsX{} & random variable\\
\Fdata {E}(\VsX{}),\Fdata {D}(\VsX{})& expectation, variance for \VsX{}\\
\Mu{\Vx{}},\Data{\Sigg{\Vx{}}} & mean, variance for \Vx{}\\
$X^\circ$ & $ X-$\STE{X} \\
$s(z)$ & Stieltjes transform\\
%

\toprule[1pt]

\end{tabularx}

\end{table}

\subsection{Situation Awareness in Power Grids}
\begin{figure}[htbp]
\centerline{
\includegraphics[width=.22\textheight,height=.3\textwidth]{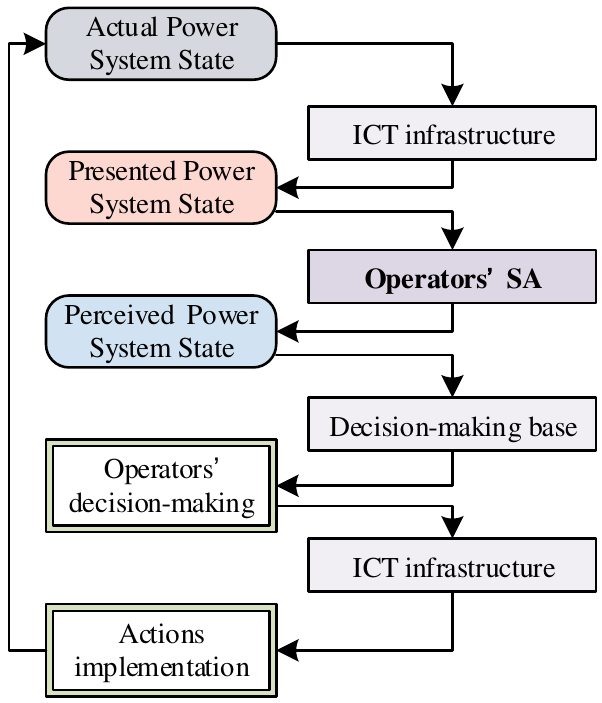}
}
\caption{SA for the operational decision-making \cite{panteli2013assessing}}
\label{fig:SA}
\end{figure}

The massive data compose the profile of the actual grid---present state;
SA aims to translate the present state into perceived state for decision-making, shown as Fig. \ref{fig:SA} \cite{he2015les, panteli2013assessing}.

In \cite{endsley2011designing}, SA is defined as  the \emph{perception} of the elements in an environment, the \emph{comprehension} of their meaning, and the \emph{projection} of their status in the near future. This paper is aimed at the use of model-free and data-driven methodology for the comprehension of the power grid.

\subsection{Load Behavior, Power Flow, and Voltage Stability}
\label{loadb}
Power flow equations define the equilibrium operating condition for a synchronous power grid. Viewed abstractly, the physical power system may be viewed as an analog power flow solution engine, taking power injection variations as \emph{inputs}, and ``computing'' bus voltage magnitude/phasors as \emph{outputs}.

Load variations, as the inputs, can be separated  into two time scales---the fast one and the slow one~\cite{lim2016svd}.
Within a short time, the fast time scale variation (time scale of \textit{seconds}) plays a dominant part; we may expect it to display small magnitude random jump behavior, filtered by the electrical characteristics of the distribution system~\cite{Brice1982PBS}---a load at a bulk distribution bus aggregates the behavior of potentially millions of individual power consuming devices, displaying on-off behavior governed by human users or local control systems. To model the fast time scale stochastic variation in a load, early studies tended to employ white noise; more recent studies have adopted Ornstein-Uhlenbeck process models~\cite{brockwell2006introduction,perninge2011modeling,lim2016svd}. We choose the white noise model for this initial study.

In the power flow equations, the four related parameters,  $\bf P,Q,$ ${\mathbf{V}} ,$ and ${\boldsymbol{\theta }}$, obey the equations~\cite{lim2016svd}:

\begin{equation}
\label{Eq:pf1}
\left[ \begin{matrix}
   \Delta \mathbf{P}  \\
   \Delta \mathbf{Q}  \\
\end{matrix} \right]
\!=\!\mathbf{J}\left[ \begin{matrix}
   \Delta\boldsymbol{\theta }  \\
   \Delta    {\mathbf{V}}  \\
\end{matrix} \right]
\!=\!\left[ \begin{matrix}
   \frac{\partial \Delta \mathbf{P}}{\partial  \boldsymbol{\theta }} & \frac{\partial \Delta \mathbf{P}}{\partial   {\mathbf{V}}  } \\
   \frac{\partial \Delta \mathbf{Q}}{\partial \boldsymbol{\theta }} & \frac{\partial \Delta \mathbf{Q}}{\partial  {\mathbf{V}}  } \\
\end{matrix} \right]\left[ \begin{matrix}
   \Delta \boldsymbol{\theta }  \\
   \Delta    {\mathbf{V}}   \\
\end{matrix} \right]\!=\!\left[ \begin{matrix}
   \mathbf{H} & \mathbf{N}  \\
   \mathbf{K} & \mathbf{L}  \\
\end{matrix} \right]\left[ \begin{matrix}
   \Delta\boldsymbol{\theta }  \\
   \Delta    {\mathbf{V}}  \\
\end{matrix} \right]
\end{equation}
then, taking the matrix inverse of the Jacobian matrix $\bf J$ in~\eqref{Eq:pf1} yields the desired input-output relationship~\eqref{Eq:pf2}:
\begin{equation}
\label{Eq:pf2}
\left[ \begin{matrix}
   \Delta \boldsymbol{\theta }  \\
   \Delta \mathbf{V}  \\
\end{matrix} \right]=\left[ \begin{matrix}
   \mathbf{M} & -\mathbf{MN}{{\mathbf{L}}^{-1}}  \\
   -{{\mathbf{L}}^{-1}}\mathbf{KM} & {{\mathbf{L}}^{-1}}+{{\mathbf{L}}^{-1}}\mathbf{KMN}{{\mathbf{L}}^{-1}}  \\
\end{matrix} \right]\left[ \begin{matrix}
   \Delta \mathbf{P}  \\
   \Delta \mathbf{Q}  \\
\end{matrix} \right]
\end{equation}
where $\mathbf{M}={{\left( \mathbf{H}-\mathbf{N}{{\mathbf{L}}^{-1}}\mathbf{K} \right)}^{-1}}$. In~\eqref{Eq:pf2}, loads injections variations $\Delta \mathbf P,\Delta \mathbf Q$ are taken as input vectors, and  $\Delta \mathbf V,  \Delta\boldsymbol{\theta } $ as output vectors.

Thus, for a certain grid where $\mathbf{Q}$ is almost constant, the model between  $\mathbf{V}$ and $\mathbf{P}$ is obtained:
  \begin{equation}
	 \Delta  {\mathbf{V}} = {\mathbf{\Xi}}\Delta {\mathbf{P}},\;\;\;{\kern 1pt} {\text{ with }}{\mathbf{\Xi}} =  - {{\mathbf{L}}^{ - 1}}{\mathbf{KM}}.
 \end{equation}
Here, the matrix $\bf \Xi$ is \textit{deterministic}, although unknown. The uncertainty of the entries in $\bf \Xi$ is neglected for mathematical tractability in this initial study. Even if $\bf \Xi$ is a random matrix, the machinery of RMT is sufficient to deal with the problem at hand.

\subsection{From Random Vector to Random Matrix Model (RMM)}
\label{sect:RMTFBD}
Considering   $T$ random vectors observed at time instants $i=1,...,T,$ we form a random matrix as follows
 \begin{equation}
\label{eq:RMTform}
	\left[ {\Delta {{{\mathbf{V}}_1}} , \cdots ,\Delta  {{{\mathbf{V}}_T}}} \right] = \left[ { {{\mathbf{\Xi}}_1\Delta{\mathbf{P}}_1}, \cdots , {{{\mathbf{\Xi}}_T\Delta \mathbf{P}}_T}} \right].
\end{equation}

In an equilibrium operating system, the voltage magnitude vector injections ${\bf V}$ with entries $V_{i},i=1,\cdots,N$ and the phase angle vector injections $\boldsymbol{\theta}$ with entries $\theta_{i},i=1,\cdots,N$ do not change a lot. Without dramatic topology changes, rich statistical empirical evidence indicates that the Jacobian matrix $\mathbf{J}$  keeps nearly constant, so does $\mathbf{\Xi}$. Also, we can estimate the changes of ${\bf V},$ $\boldsymbol{\theta},$ and $\mathbf{\Xi}$ only with the \emph{empirical} approach.
Thus  we rewrite \eqref{eq:RMTform} as:

\begin{equation}
\label{Eq:RMMVTP}
\mathbb{V} = {\MatrxS {\Xi}N}\Matrx {\mathbb{P}}NT
\end{equation}
where $\mathbb{V} = \left[ {\Delta {{{\mathbf{V}}_1}} , \cdots ,\Delta  {{{\mathbf{V}}_T}}} \right]$, $ {\MatrxS {\Xi}{}}={\mathbf{\Xi}}_1=\cdots={\mathbf{\Xi}}_T,$ and $\mathbb{P} = \left[ {\Delta {{{\mathbf{P}}_1}} , \cdots ,\Delta  {{{\mathbf{P}}_T}}} \right].$
Here $\mathbb{V}$ and $\mathbb{P}$ are random matrices. In particular, $\mathbb{P}$ is a random matrix with Gaussian random variables as its entries.

\subsection{Hypothesis Testing for Anomaly Detection}
One of the common ways to perform data-driven anomaly detection is to employ statistical models which are usually trained out using data from \emph{faulty and nominal behavior} in a batch mode. However, it is difficult to anticipate, \emph{a priori}, all the possible ways in which failures may occur, especially when a new vehicle model is introduced \cite{sankavaram2015incremental}.
As an alternative, based on RMM we can build the statistical models only using sampling data and employ hypothesis testing for anomaly detection. Moreover, together with methods/tools such as moving split window \cite{he2015arch}, augmented matrix \cite{he2015corr}, we design SA functions such as fault isolation/classification/diagnosis.

For a system described as \emph{\ref{sect:RMTFBD}}, we assume that we sample fast enough and there are no line/unit tripping events\footnote{See section \emph{\ref{Intro}}, examples of the events include changes of network topology, dramatic increases of power demand at some bus, and so on.} occurring, according to the arguments in \emph{\ref{loadb}}. With the above assumptions,  $\Delta {\mathbf{P}_i}$ is, as a result, modeled as Gaussian random vectors, as done in~\cite{lim2016svd}.
Then, as \eqref{Eq:RMMVTP}, the Gaussian random matrix $\mathbb{P}_{N\times{T}}$ is employed to model the system power injections, and the corresponding $\mathbb{V}$ is obtained.
The $\mathbb{V}$ is turned $\tilde{\mathbb{V}}$ with a normalization procedure in  \cite{he2015arch}.
For the normalized matrix $\tilde{\mathbb{V}}$, we formulate our problem of anomaly detection in terms of the binary hypothesis testing:  no event exists $ {\cal H}_0 $, and signal exists $ {\cal H}_1$:
\begin{equation}\label{eq:ADtest11} 
\left| \begin{aligned}
  & {{\mathcal{H}}_{0}}:\tilde{\mathbb{V}} = \tilde{\mathbf{\Xi}}\Matrx RNT \\
 & {{\mathcal{H}}_{1}}:\tilde{\mathbb{V}} \neq \tilde{\mathbf{\Xi}}\Matrx RNT \\
\end{aligned} \right.
\end{equation}
where $\mathbf{R}$ is the standard Gaussian random matrix, i.e., its entries $r_{i,j}$ are independent identically distributed (i.i.d.) Gaussian.
The sample covariance matrix $\mathbf{S}$ is defined as:
\begin{equation}
\label{Eq:SCM}
\mathbf{S} =\frac{1}{T}\mathbf{X}\mathbf{X}^H
\end{equation}
For $\mathbf{R}$, a special case, its sample covariance matrix is
\begin{equation}
\label{Eq:SCMRR}
\mathbf{S}_{0} =\frac{1}{T}\mathbf{R}\mathbf{R}^H
\end{equation}
If signals (see \emph{\ref{ContriNS}}) exist in the system, ${\mathcal{H}}_{0}$ will be reject.

\subsection{RMM Analysis}
\label{RMMAna}
Consider the sample covariance matrix of  $\tilde{\mathbb{V}}$ defined in~\eqref{eq:ADtest11} under  $ {\cal H}_0 $.:
\begin{equation}
\label{Eq:RMMNorm11}
\underline{\mathbf{B}}_N=\frac{1}{T}\tilde{\mathbb{V}}\tilde{\mathbb{V}}^H=\frac{1}{T}\tilde{\mathbf{\Xi}}\mathbf{R}\mathbf{R}^H\tilde{\mathbf{\Xi}}^H
=\tilde{\mathbf{\Xi}}\mathbf{S}\tilde{\mathbf{\Xi}}^H
\end{equation}

Firstly, we define the following notations \cite{yao2015large}:
\begin{equation}
\label {eq:SampleCN}
\underline{\mathbf{C}}_{N}=\mathbf{S}_0 \mathbf{\mathbf{T}}
\end{equation}
\begin{equation}
\label {eq:SampleDN}
\underline{\mathbf{D}}_{N}={{\mathbf{T}^{1/2}}}\mathbf{S}_{0}\mathbf{{\mathbf{T}^{1/2}}}
\end{equation}
\begin{equation}
\label {eq:SampleCNN}
\overline{\mathbf{C}}_{T}=\frac{1}{T}\mathbf{R}{^H}\mathbf{T}\mathbf{R}
\end{equation}
where ${\mathbf{T}}= {\tilde{{\mathbf{\Xi}}}^H}\tilde{{\mathbf{\Xi}}}$ is a non-negative Hermitian matrix; ${\mathbf{T }}$ is deterministic and independent of $\mathbf{S}_0$. $\underline{\mathbf{B}}$, $\underline{\mathbf{C}}$,  $\underline{\mathbf{D}}$, and $\overline{\mathbf{C}}$ share the same non-null eigenvalues \cite{yao2015large}. We need this fact to use Lemma~\ref{Lemma1}.

\newtheorem{thm4}{Lemma}[section]
\begin{thm4}[2010, \cite{Bai2010BookRMT},  Section 2.4]
\label{Lemma1}
Let $\mathbf{S}_n$ be the sample covariance matrix defined in \eqref{Eq:SCM} with i.i.d. components and let $\mathbf{T}_n$ be a sequence of nonnegative definite Hermitian matrices of size $p\times{p}.$ Define  $\mathbf{B}_n=\mathbf{S}_n\mathbf{T}_n$ and assume the following.
\begin{itemize}
\item Entries $x_{i,j}$ of  $\mathbf{X}$ are i.i.d. with mean $0$ and variance $1$.
\item $\mathbf{X}$'s sample size ratio $p/n\rightarrow{y}>0$ when $n\rightarrow \inf$.
\item $\mathbf{T}_n$ is either deterministic or independent of $\mathbf{S}_n$.
\item Almost surely, the sequence $H_n=F^{\mathbf{T}_n}$ of the empirical spectral distribution (ESD) of ${\mathbf{T}_n}$ weakly converges to a nonrandom probability measure $H$.
\end{itemize}
Then almost surely, $F^{\mathbf{B}_n}$ weakly converges to a nonrandom probability measure $F_{y,H}.$ Moreover its Stieltjes transform $s(z)$ is implicitly defined by the equation
\begin{equation}
\label {eq:M_PEquation}
s\left( z \right)=\int{\frac{1}{t\left( 1-y-yzs\left( z \right) \right)-z}\text{d}}H\left( t \right),z\in {{\mathbb{C}}^{+}}
\end{equation}
\end{thm4}

Comparing with \eqref{eq:SampleCN} indicates that \eqref{eq:SampleCN} complies all the assumptions in Lemma \ref{Lemma1},  by noting that the entries of $\mathbf{X}$ are Gaussian random variables, thus a special case of i.i.d..

The standard M-P distribution is easily recovered from \eqref{eq:M_PEquation}. In this case, $\mathbf{T}_n=\mathbf{I}_p$ so that $H=\delta_1$, and \eqref{eq:M_PEquation} becomes:
\begin{equation}
\label {eq:standMP}
s\left( z \right)={\frac{1}{ 1-y-yzs\left( z \right)-z}}
\end{equation}

Consider another form, a $p \times n$ matrix
\begin{equation}
	{{\mathbf{Y}}_n} = \frac{1}{{\sqrt n }}{\mathbf{\Sigma }}_n^{1/2}{{\mathbf{X}}_n},
\end{equation}
where ${{\mathbf{\Sigma }}_n}$ is a nonnegative definite Hermitian matrix and ${\bf X}_n$  is a random matrix with i.i.d. real or complex standardized entries. The fluctuations of the LES \cite{lytova2009clrforles}
\begin{equation}
\label {eq:LESYYYY}
\operatorname{Tr} \varphi \left( {{{\mathbf{Y}}_n}{\mathbf{Y}}_n^H} \right) = \sum\limits_{i = 1}^p {\varphi \left( {{\lambda _i}} \right)} ,\;\;\;{\kern 1pt} {\lambda _i} = {\text{eigenvalues of }}{{\mathbf{Y}}_n}{\mathbf{Y}}_n^H
\end{equation}
are shown in~\cite{najim2013gaussian} to be Gaussian, in the regime that both the dimensions  $p$ and $n$  goes to infinity at the same pace and in the case where $\varphi$ is an analytical function.
Note that
\begin{equation}
\label {eq:YYYY}
\begin{aligned}
\operatorname{Tr} \varphi \left( {{{\mathbf{Y}}_n}{\mathbf{Y}}_n^H} \right)& =
\operatorname{Tr} \varphi \left( \frac{1}{{\sqrt n }} \mathbf{\Sigma }_n^{1/2}{{\mathbf{X}}_n}{{\mathbf{X}}_n}^H{\mathbf{\Sigma }}_n^{1/2} \right)\\
&=\operatorname{Tr}\varphi \left( \mathbf{\Sigma}_n^{1/2} \mathbf{S}_n \mathbf{\Sigma}_n^{1/2} \right)=\operatorname{Tr} \varphi \left(  \mathbf{S}_n \mathbf{\Sigma}_n \right)
\end{aligned}
\end{equation}

As a result, we build the \emph{connection} between large random matrices and big data analytics.
Fully driven by sampling data, we can obtain the experimental LESs via \eqref{eq:LESYYYY}. Note that the $\mathbf {Y}$ in \eqref{eq:LESYYYY} is tied together with the sampling raw data $ {\mathbb V}$
through \eqref{eq:ADtest11}, \eqref{Eq:RMMNorm11}, \eqref{eq:SampleCN}, and \eqref{eq:YYYY}.
We see~\cite{Qiu2016BigDataLRM} for more details about the mathematical foundations.

\section{Mathematical Statistics on Large Random Matrices}
\label{section: backgroundMath}

\subsection{From Physical Grid to Random Matrix Model}
\label{GridToRMM}
For a power grid, there exist various parameters (e.g., frequency, voltage, current, and power). Among these, we often choose bus voltage magnitudes for further analysis.
According to \eqref{eq:RMTform}, we can obtain a random data matrix denoted as  \Matrx{\VOG{}}{n}{t} with \Data {t} observations for a \Data {n}-dimensional random vector.

The \Matrx{\VOG{}}{n}{t} is in a high-dimensional space but not an infinite one; more explicitly, we are interested in the practical regime for a system of \Equs{\Data {n}}{}{100--3000} nodes.
Any temporal and spacial sections can be considered in  \VOG{}, e.g., selecting $T$ data points from $t$, and $N$ nodes from $n$. These data points ${\Rdata v_{i,j}}$  form a matrix, denoted as \Matrx {\hat{\mathbf{X}}}{N}{T}.
Note that \Matrx{\VOG{}}{n}{t} can be updated as time goes by in a manner that the last column is the nearly real-time data.

The {\VRX {}} is a random matrix due to the presence of ubiquitous uncertainties in the system. Furthermore, we can convert \VRX {} into a normalized matrix {\VTX{}}  row-by-row:
\FuncC {eq:StdMatrix}
{
\VTx{i}\!=\!\Div{1}{\sigma_i({\VRx{i}})}{(\VRx{i}\!-\!\mu_i( {\VRx{i}})}),\quad\Data {1\!\LE{}\!i\LE{}\!N\!}
}
where \Equs {\VRx{i}}{(\hat x_{i,1},\!\cdots\!,\hat x_{i,T})}.  The $\mu_i$ denotes the mean for every node $i=1,\!\cdots\!,N,$ and  the $\sigma_i$ denotes the standard deviation.  Then we can conduct RMM analysis according to \emph{\ref{RMMAna}}.

\subsection{Random Matrix Theory (RMT)}
\normalsize{}
The mathematical tools of RMT include the Ring Law, the M-P Law, and the MSR.

\subsubsection{Single Ring Theorem}
{\Text{\\}}

The Single Ring Theorem~\cite{guionnet2011single, tao2013random} describes the ESD of a large generic matrix  with prescribed singular values, i.e. an \Muls{N}{N} matrix of the form $\mathbf{A}\!=\!\mathbf{U}_1 \mathbf{\Upsilon} \mathbf{U}_2$, with $\mathbf{U}_1$ and $\mathbf{U}_2$ some independent Haar-distributed unitary matrices and  $\mathbf{\Upsilon}$ a deterministic matrix whose singular values are the ones prescribed. 
\subsubsection{Marchenko-Pastur Law (M-P Law)}
{\Text{\\}}
In \emph{\ref{sect:RMTFBD}}, we show that the ESD follows the generalized M-P distribution and satisfies the condition of being compactly supported limit measure. As a result, the Single Ring Theorem says that the empirical distribution is a single ring on the complex plane.

\newtheorem{thm1}{Lemma}[section]
\begin{thm1}[Marchenko-Pastur Law, 1967, \cite{marvcenko1967distribution}]
Assume a \Muls {N}{T} random  matrix \VX{} obeys condition\footnote{The matrix ensemble is said to obey condition $\mathbf{C1}$ with constant  $C_0$ if the random variables $X_{i,j}$ are jointly independently, having zero mean and variance one, and obey the moment condition $\underset{i,j}{\mathop{\sup }}\,\mathbb{E}{{\left| {{X}_{i,j}} \right|}^{{{C}_{0}}}}\le C$.} $\mathbf{C1}$ with $C_0\ge4$, and $N,T\!\to\! \infty$ such that $\underset{N\to \infty }{\mathop{\lim }}\,\!N\!/\!T\!=\!c\!\in (0,1]$, the \text{ESD} of the matrix  {\Vector S} defined in \eqref{Eq:SCM} converges in distribution to the M-P Law with a density function:
\FuncCC {eq:MPLaw}
{\Funcfzx {\rho}{\text{mp}}{\Vlambda{}}}
{ \Div{1}{2\pi{}\lambda c{\sigma^2}}\Sqrt{ (a_+-\lambda)(\lambda-a_-)} }{{\Text {, }} a_-\LE \lambda \LE a_+ }
{0}  {\Text {, otherwise}}
{where \Equs {a_\pm}{\sigma^2(1\pm\Sqrt c)^2}.}\normalsize{}
\end{thm1}

For the special case of i.i.d., see \emph{\ref{caseiid}}.

\subsubsection{Universality Principle}
{\Text{\\}}

This principle~\cite{vanprobability} enables us to obtain the exact asymptotic distributions of various test statistics without restrictive distributional assumptions of matrix entries~\cite{qiu2013bookcogsen}. For a real system, we cannot expect the matrix entries to follow i.i.d. distribution. One can perform various hypothesis tests under the assumption that the matrix entries are not Gaussian distributed but use the same test statistics as in the Gaussian case.  Numerous studies using both simulations  and experiments demonstrate that the Ring Law and  the M-P Law are universally valid---the asymptotic results are remarkably accurate for relatively moderate matrix sizes such as tens~\cite{he2015arch}. This is the very reason why RMT can handle practical massive systems.

\subsubsection{Independent Identically Distributed Case}
\label{caseiid}
{\Text{\\}}

For a rectangular \Muls {N}{T} random matrix \VX{}, the entries are i.i.d. variables, satisfying
 \begin{equation}
 \label{eq:randomX}
 \Equs {\STE{\VsX{i,j}}}{0}, \quad  \Equs {\STE{{\VsX{i,j}}{\VsX{m,n}}}}{\delta_{i,m}\delta_{j,n}{\sigma^2}}
 \end{equation}
 {}where ${\delta}$ is the Kronecker Delta Function defined as
\[
\delta_{\alpha,\beta}=
\begin{cases}
    1 & \alpha=\beta \\
    0 & \alpha\neq\beta.
\end{cases}
\]

For the normalized data matrix \VTX{} obtained as \eqref{eq:StdMatrix}, according to~\cite{ipsen2014weak}, we define its singular value equivalent as
\[
{{\Equ {\VX u}   {\Sqrt{\VTX{}{\Her{\VTX{}}}}\Vector U }} },  \text{ \Belong {\Vector U} {\VF CNN} is a Haar unitary matrix.}
\]

Then, for $L$ independent matrices \VX {u,i}, we consider their product
\[\Data {\VZ{}=\Prod{i=1}{L} \VX {u,i}}.\]
The product \VZ{} is converted into \VTZ {} as
\[
{\Equ {\VTz{i}} {\Vz{i}/({\Sqrt N}\sigma_i({\Vz{i}}))}},
\quad \Data {1\LE{}i\LE{}N}
\]
{where \Equs {\Vz{i}}{(z_{i,1},\cdots,z_{i,N}).}\normalsize{}
  Based on \eqref{eq:SRab} and \eqref{eq:MPLaw}, the ESD of \VTZ {} converges almost surely to the same limit given by \cite{qiu2015smart}
\FuncCC {eq:RingLaw}
{\Funcfzx {\rho}{\text{ring}}{\Vlambda {}}}
{ \Div{1}{\pi{}c\Data {L}}{\Power{\Abs{\lambda}}{(2/\Data {L}-2)}}}  {{\Text {, }} \Power {(1-c)}{\Data {L}/2} \LE \Abs{\lambda} \LE \Data 1   }
{0}   {\Text {, otherwise}}
as \Trend {N,T}{\INF} with the ratio \Equ {c}{\Belong {N/T} {(0,1]} }.

\subsubsection{Mean Spectral Radius (MSR)}
{\Text{\\}}

The geometrically motivated MSR is a statistic that is a complicated function of collected data vectors;  it may also be viewed as  a certain LES, as defined in~\eqref{eq:LESYYYY}. For any matrix, such as \VTZ{}, we can obtain its eigenvalues ${\lambda}_i(\VTZ{}),i=1,...,N$ on the complex plane. The mean value of all these eigenvalues' radii  is denoted as  {\VLES{MSR}:
\FuncC {eq:MSR}{
\Equs {\VLES{MSR}{}}{ {\Div{1}{N}\Sum {i=1}{N}\Abs {\lambda_i(\VTZ{})}}, }}
}
where $N$ is typically large, say in the order of $100-3000.$ The large value of $N$ has a far-reaching effect. That is the reason why the Gaussian properties of such LESs as \VLES{MSR} may be justified~\cite{Bai2010BookRMT,Qiu2016BigDataLRM}. The facts are analogues of the CLTs in classical probability.

\section{Linear Eigenvalue Statistics of Large Random Matrices}
\label{section: Les}
\subsection{Linear Eigenvalue Statistics}
\subsubsection{Definition}
{\Text{\\}}

The LES $\tau$ of an arbitrary matrix \Belong{\mathbf{\Gamma}}{\VF CNN} is defined via the continuous test function  \Data{\varphi: \Fdata C \rightarrow \Fdata C.}  \cite{lytova2009clrforles,shcherbina2011central}

\begin{equation}
\label{eq:DDLES}
\tau(\varphi, \mathbf{\Gamma})=\mathcal{N}_N[\varphi]=\Sum{i=1}{N}{\varphi({\lambda_i})}=\text{Tr}\varphi \left( \mathbf{\Gamma} \right)
\end{equation}
where the trace of the function of a random matrix is involved.

It is very interesting to study the special case for a Gaussian random matrix. For power grids, the universality principle is extremely relevant in this context.

\subsubsection{Law of Large Numbers}
{\Text{\\}}

The law of Large Numbers tells us that $N^{-1}\mathcal{N}_N[\varphi]$ converges in probability to the limit
\begin{equation}
\label{eq:LES1}
\lim_{N \to \INF}\Div 1N\mathcal{N}_N[\varphi]\!=\!\int\varphi(\lambda)\rho(\lambda)\,d\lambda
\end{equation}
{where $\rho(\lambda)$ is the probability density function of $\lambda$.}\normalsize{}
%

\subsubsection{Central Limit Theorem of LES}
{\Text{\\}}

The CLT \cite{johansson1998fluctuations,guionnet2002large,bai2004clt,anderson2006clt,lytova2009clrforles,shcherbina2011central,pan2011universality}, as the natural second step, aims to study the LES fluctuations. Consider another form of covariance matrix:
\begin{equation}
\label {eq:Sc2}
\Equs {{\Vector M}}{\Div 1 N\VX{}\Her{\VX{}}}=\Div {1}{c}\Vector S_N
\end{equation}
According to \eqref{eq:MPLaw}, its ESD converges to
\FuncCC {eq:MPLaw2}
{\Funcfzx {\rho}{\Data{\!m\!p}_2}{\Vlambda{}}}
{ \Div{1}{2\pi{}\lambda {\sigma^2}}\Sqrt{ (b_+-\lambda)(\lambda-b_-)} }{{\Text {, }} b_-\LE \lambda \LE b_+ }
{0}  {\Text {, otherwise}}
{where \Equs {b_\pm}{\sigma^2(1\pm1/\Sqrt c)^2}.}\normalsize{}

The CLT for {\Vector M} is given as follows \cite{shcherbina2011central}:
\normalsize{}

\newtheorem{thm2}{Theorem}[section]
\begin{thm2}[M. Sheherbina, 2009]
 Let the real valued test function $\varphi$ satisfy condition ${{\left\| \varphi  \right\|}_{3/2+\varepsilon }}<\infty   \left( \varepsilon >0 \right)$. Then ${\mathcal{N}_N}^\circ[\varphi]$ defined in \eqref{eq:LES1}, in the limit $N,T\to \infty , \Equs {c}{N/T}\le 1$, converges in the distribution to the Gaussian random variable with zero mean and the variance:
{}
\begin{equation}
\label {eq:CLTforLes}
\begin{aligned}
     {{V}_{SC}}\left[ \varphi  \right]=    &    \frac{2}{c\pi^2 }\iint\limits_{-\frac{\pi }{2}<{{\theta }_{1}},{{\theta }_{2}}<\frac{\pi }{2}}{{{\psi }^{2}}\left( {{\theta }_{1}},{{\theta }_{2}} \right)}\left( 1-\sin {{\theta }_{1}}\sin {{\theta }_{2}} \right) d{{\theta }_{1}}d{{\theta }_{2}} \\
 &                +\frac{{{\kappa}_{4}}}{{\pi }^{2}}\left( \int_{-\frac{\pi }{2}}^{\frac{\pi }{2}}{\varphi \left( \zeta \left( \theta  \right) \right)\sin \theta  d{{\theta }}} \right)^2 \\
\end{aligned}
\end{equation}
{where $\psi \left( {{\theta }_{1}},{{\theta }_{2}} \right)\!=\!\frac{\left[ \varphi \left( \zeta \left( \theta  \right) \right) \right]\arrowvert_{\theta ={{\theta }_{2}}}^{\theta ={{\theta }_{1}}}}{\left[ \zeta \left( \theta  \right) \right]\arrowvert_{\theta ={{\theta }_{2}}}^{\theta ={{\theta }_{1}}}},$
${\left[ {\zeta \left( \theta  \right)} \right]\arrowvert_{\theta  = {\theta _2}}^{\theta  = {\theta _1}}}\!=\!\zeta \left( {{\theta _1}} \right)\! -\! \zeta \left( {{\theta _2}} \right),$
and $\zeta \left( \theta  \right) \!= \!1\! + \!1/c \!+\! {2}\!/\!{\sqrt c} \sin \theta;$   $\kappa_4\!=\!\mathbb{E}\left( {{X}^{4}} \right)\! -\!3$ is the $4$-th cumulant of entries of \VX{}.}

\normalsize{}
\end{thm2}
\normalsize{}

\subsection{LES Designs and the Theoretical Values}
\label{LESdesign}

A typical scenario is assumed to obtain the theoretical values:
\Equs {N}{118}, \Equs {T}{240}, and \Equs {c}{N/T}\Equs {}{0.4917}.

\subsubsection{LES for Ring Law}
\Text{\\}

The MSR defined as  \eqref{eq:MSR} is in a special form of a LES. Since $\lambda_i({\widetilde{\mathbf Z}}),i=1,...,N$ are highly correlated random variables; each one is a complicated function of the raw random matrices \VRX{}, and thus $\tau_{\text{MSR}}$ itself is a random variable, the sum of dependant random variables divided by the total number. The self-averaging property~\cite{Qiu2016BigDataLRM} with a large matrix size $N$ is remarkable.
According to \eqref{eq:RingLaw} and \eqref{eq:LES1}, the theoretical expectation of the random variable $\tau_{\text{MSR}}$, as $N \to \INF,$ approaches asymptotically the limit:
\begin{equation}\label{eq:ExpMSR}
\begin{aligned}
\STE{\VLES{MSR}{}}
&\Data{= \iint_{\text {Area}}\rho_{\text{ring}}(r){\times}r{\cdot}r\,{\rm d}r   {\rm d}\theta}\\
&\Data{= \int^{2\pi}_0\!\int^1_{\sqrt{1-c}}   \Div{1}{c\pi}r{\cdot}r  {\rm d}r \;  {\rm d}\theta}=0.8645\\
\end{aligned}
\end{equation}
In practice, we can use this asymptotic limit to replace the true value (unknown) for a finite value of $N,$ when $N$ is sufficiently large.

\subsubsection{LES for Covariance Matrices}
\Text{\\}

1. Chebyshev Polynomials ($T_2$): ${\varphi(\lambda)=2x^2-1}$
\begin{equation}\label{eq:LEST2}
\tau_{T_2}=\sum_{i=1}^N(2{\lambda_i}^2-1),
\end{equation}
and according to \eqref{eq:DDLES}, \eqref{eq:LES1}, and \eqref{eq:MPLaw2}, we obtain
\begin{equation}\label{eq:TheE}
\begin{aligned}
\STE{\tau_{T_2}}  &=N\!\int{\varphi(\lambda)}\rho_{\text{mp2}}(\lambda)\,d\lambda=1338.3.\\
\end{aligned}
\end{equation}
 According to \eqref{eq:CLTforLes}, we write the variance as
\begin{equation}\label{eq:TheD}
\STD{\tau_{T_2}}=665.26.
\end{equation}
Similarly, for a rectangular data matrix $\Gamma$,  according to \eqref{eq:DDLES}, we can obtain various LESs $\tau(\varphi, \mathbf{\Gamma})$ via  designing different test functions,  as well as their theoretical values. Here we list some classical test functions:

2. Chebyshev Polynomials ($T_3$): $ \varphi(x)=4x^3-3x$

3. Chebyshev Polynomials ($T_4$): $ \varphi(x)=8x^4-8x^2+1$

4. Generalized variance or determinant (DET): $\varphi(x)=\ln(x)$

5. Likelihood-ratio test (LRT): $\varphi(x)=x-\ln(x)-1$
\subsection{LES Indicator and Classical One}
The methodology of SA based on a class of LESs, including technical route, concrete procedures, evaluation system, and advantages over classical one, is another meaningful topic. Here, we do not expand on this and  just show the characteristics from the indicator perspective.

According to \emph{\ref{LESdesign}}, numerous LESs are able to be designed for a ceratin RMM \VX{}.
With these LESs, the high-dimensional indicator system is built; it gives us a multiple view angle to gain insight into the system---the test function is similar to a filter in some sense.
Table \ref{tab:Indicator} lists the features of the LES indicator and makes a comparison with the classical one.

\begin{table*}[htbp]
\caption{Indicator System for Power System}
\label{tab:Indicator}
\centering
\begin{tabularx}{1\textwidth} { p{8.6cm} !{\color{black}\vrule width1pt} p{8.6cm} } 
\toprule[1.5pt]
\hline

\textit{LES Indicator} & \textit{Classical Indicator}\\
\Xhline{1pt}
data-driven and model free & model based\\
supported by novel theorems& supported by tradition laws\\
unclearly defined engineering concept & clearly defined one\\
sensors and network of high quality are required & sensors and network are OK is enough\\
\hline
probabilistic value& accurate one\\
statistic in high dimensions & often in low dimensions\\
the value relies on massive (all) data & the value relies on a few data ($<$model's dimensions)\\
robust against bad data and insensitive to individual data& sensitive to sample selections and the training\\
\hline
pure statistical procedure without system errors&system errors are inevitable\\
random errors can be estimated with the model size ($N,T$)& the errors depend on the model building procedure\\
\hline
compatible with diverse data& only for the assigned ones (relying on the model)\\
readily transformation in statistical space (preprocessing)& in a fix/inflexible form\\
naturally decoupling the interconnected system & decoupling based on assumptions and simplifications\\
\hline
\toprule[1pt]
\end{tabularx}
\end{table*}

As pointed out in Table \ref{tab:Indicator}, the LES indicator system provides a much better way, compared to the classical one, to utilize the massive data for a complex system.
The relation of the former to the later, in some sense, is just like that of the  quantum physics\footnote{The quantum physics is one of the greatest achievements in the history of physics; it is unreasonably effective and still influences and puzzles us deeply nowadays \cite{cao2008quantum}.} to the classical one.
Instead of accurate measurements and descriptions (for the mass, position, velocity, etc), the later gives probabilistic statistics. By comparing the experimental values (always fully data-driven) with ideal theoretical values (guaranteed by theorems), the complicated system is understood statistically.

In short,  with the help of RMT, we can investigate a data-driven methodology, based on which we can turn the real power grid into RMMs, using measurements such as PMUs. With a pure mathematical procedure, a class of  statistical indicators is formed as a new epistemology for the grid. Some advantages of this epistemology---such as data-driven and model-free, speed, reasonableness, sensitivity---have already been shown in our previous work \cite{he2015arch, he2015corr}; we will show its flexibility and robustness against bad data in the \emph{Section \ref{section: case}}.

\section{Functions Design with LES and Related Cases}
\label{section: case}
\normalsize{}
For the background, we adopt  IEEE 118-bus system with 6 partitions \cite{he2015arch} (Fig. \ref{fig:IEEE118network}). The SA functions, \emph{anomaly detection} and \emph{fault diagnosis}, and \emph{correlation analysis} based on MSR, have already been designed in our previous work \cite{he2015arch} and \cite{he2015corr} respectively. This paper pushes the design forward.

\subsection{SA based on Theoretical Exception}
We assume the events as shown in Table \ref{Tab: Event Series}; the final result is illustrated as the solid blue line in Fig. \ref{fig:MSR1}.
According to 
\emph{\ref{GridToRMM}},
we obtain a data source $\mathbf{\Omega}_{\mathbf{V} }:{\Rdata v_{i,j}}\Belong{}{\VF R{118}{1600}}$ for $N=118$ nodes with $T=1600$ time instants.
Noting that the grid fluctuations are set by \Vgam{\Text {Acc}} and \Vgam{\Text {Mul}} as \cite{he2015arch}
\FuncC{eq:gridfluctuation}{
\Equs {\Index {\Tdata y}{\Text{load\_}nt}}  {
\Index  y{\Text{load\_}nt}\Mul{}{}(1+\Muls{\Vgam{\Text {Mul}}}{r_1})
} +   \Muls{\Vgam{\Text {Acc}}}{r_2}
} 		
where $r_1$ and $r_2$ are standard Gaussian random variable.

\begin{table}[H]
\caption {Series of Events}
\label{Tab: Event Series}
\centering

\begin{minipage}[htbp]{0.48\textwidth}
\centering

\begin{tabularx}{\textwidth} { >{\scshape}l !{\color{black}\vrule width1pt}    >{$}l<{$}|    >{$}l<{$}|   >{$}l<{$}| >{$}l<{$} }  
\toprule[1.5pt]
\hline
\textbf {Stage} & \textbf{E1} & \textbf{E2} & \textbf{E3} & \textbf{E4}\\
\toprule[1pt]
Time (s) & $1--400$ & $401--800$ & $801--1200$ & $1201--1600$ \\
\hline
\VPbus{52} (MW) & 18 & 18  & 300 & t/3-100\\
\hline
$P_{Fluctuation}$ & \text{none} & \multicolumn{3}{l} {small: \Vgam{\Text {Acc}}\!=0.1, \Vgam{\Text {Mul}}\!=0.001} \\
\toprule[1pt]
\end{tabularx}
\raggedright
\small {*\VPbus{52} is power demand of bus-52.
}.
\end{minipage}
\end{table}

With the previous work \cite{he2015arch}, the \Cur{\tau_{\text{MSR}}}{t} curve is obtained, as illustrated in Fig. \ref{fig:MSR1}, to conduct anomaly detection. The $\tau_{\text{MSR}}$ starts the dramatic changes at the instants \Et{801}{s} and \Et{1378}{s}; these are right the beginning for the event of step-change and voltage collapse, respectively.

\begin{figure}[htbp]
\centering

\includegraphics[width=0.48\textwidth]{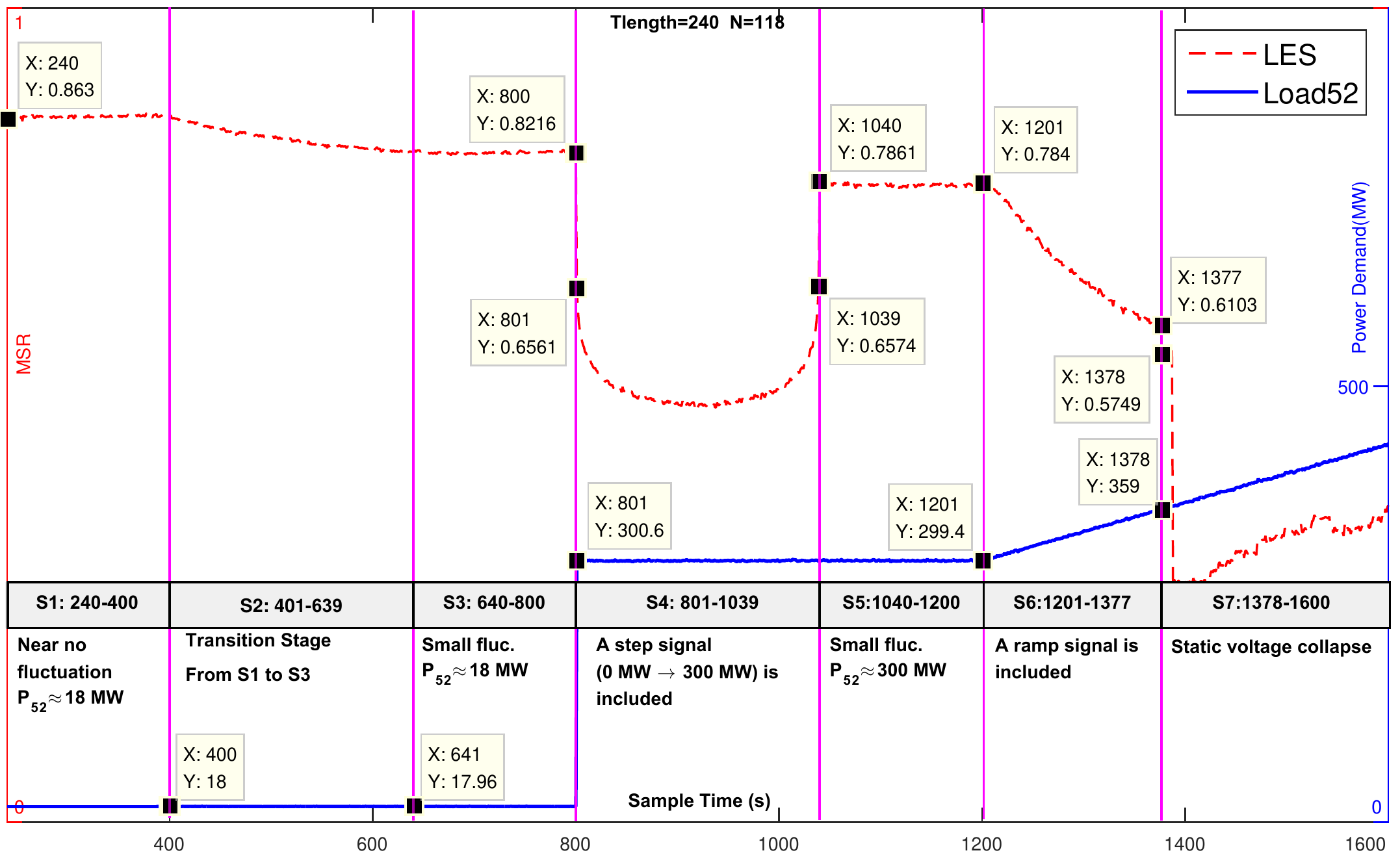}

\caption{\Cur{\tau_{\text{MSR}}{t}} curve and stage division}
\label{fig:MSR1}
\end{figure}

\subsubsection{LES Designs and Static Stability Evaluation}
\Text{\\}

Keep \Equs{N}{118} and \Equs{T}{240}. According to \emph {\ref{LESdesign}}, numerous LESs $\tau$ are designed  and they empirically  converge in the distribution to the Gaussian variables with   means $\Fdata {E}({\tau})$ and  variances $\Fdata {D}({\tau})$. These theoretical values are able to be obtained via a similar statistical process to \eqref{eq:ExpMSR}, \eqref{eq:LEST2}, \eqref{eq:TheE}, and \eqref{eq:TheD}, and are shown in  \textbf{E0} part of Table  \ref{Tab: LES}.

On the other side, the experimental LESs at each time instant $t$, denoted as $\tau_{\varphi}(t)$, are obtained via a \emph{fully data driven} procedure, in turn, as  \eqref{eq:StdMatrix}, \eqref{eq:Sc2} and \eqref{eq:DDLES}.
Duration a stage, we acquire $\tau_{\varphi}(t_1),\!\cdots\!,\tau_{\varphi}(t_k)$, and then, their  statistics---Mean \Mu{\tau} and Variance \Sig{\tau}. Table  \ref{Tab: LES} is mainly such an illustration of these statistics for the given case.
Taking the raw data about voltage magnitude as an example, following steps are needed as the  preprocess:
\begin{itemize}
\item We can set the data-processing interval, i.e. the step between $\tau_{\varphi}(t_j)$ and $\tau_{\varphi}( t_{j+1})$,  to a single sampling time point (i.e. $\Delta t_{\text{DP}}\!=\!1$) to conduct one-by-one moving split-window analysis, or to $T$ points (i.e. $\Delta t_{\text{DP}}\!=\!T$) to make $\tau_{\varphi}(t_1),\!\cdots\!,\tau_{\varphi}(t_k)$  independent\footnote{They do not shall a same vector.} on the time series.
\item There exist, in a system, numerous $PV$ nodes and $V\theta$ nodes whose voltage magnitudes keep constant and will lead to a singular matrix; there are also some small signals, causing by input noises, needed to be filtered. Therefore, we  add a small random fluctuation; the signal-to-noise ratio (SNR) is set  as 0.002 experientially, i.e., $v_{ij}\!=\!v_{ij}\!+\!0.002r$, where $r$ is a Gaussian random variable.
\item The normalization  is introduced as \eqref{eq:StdMatrix}.
\end{itemize}

Above steps will change the theoretical variation to some value, denoted as $\Fdata {D}_{\text{1}}({\tau})$; $\Fdata {D}_{\text{1}}({\tau})$ is able to be obtained in advance by simulations using the ideal inputs, i.e., standard random matrix \Matrx {X}{118}{240}, with the same data preprocessing.

Suppose the sampling frequency is high enough and the $\Delta t_{\text{DP}}\!=\!T$. Thus, for a constant state, such as  \textbf{E1}: \emph{no fluctuations around 18 MW}, we can obtain more than $40$ independent RMMs $\Belong{\VX{i}}{\VF R{118}{240}}\  (i=1,\cdots,40,\cdots)$.
\textbf{E1}--\textbf{E3} parts show these experimental statistics; sorted by the deviations of the means from its theoretical predictions, it goes that $\textbf{E1}<\textbf{E2}<\textbf{E3}$. 

\subsubsection{LES System Performance for SA in this Case}
\Text{\\}

Back to the assumed scenes, according to the anomaly source \VPbus{52} and window size $T$, we divide the temporal space into 7 stages: \textbf{S1}--\textbf{S7}, shown as Fig. \ref{fig:MSR1}. Parts \textbf{S1}--\textbf{S7} in Table \ref{Tab: LES} show the performance of the designed LESs. Roughly speaking,  sorted by  the means they are arranged as $\textbf{S1}<\textbf{S2}<\textbf{S3}<\textbf{S5}<\textbf{S6}<<\textbf{S4}<<\textbf{S7}$, and by the variances as $\max(\textbf{S1, S3, S5})<<\textbf{S2}<<\textbf{S4}<<\textbf{S6}<<\textbf{S7}$.

\begin{table}[htbp]
\caption {LESs and their Values}
\label{Tab: LES}
\centering

\begin{minipage}[!h]{0.49\textwidth}
\centering

\footnotesize
\begin{tabularx}{\textwidth} { >{\scshape}l !{\color{black}\vrule width1pt}        >{$}l<{$}  >{$}l<{$}  >{$}l<{$}   >{$}l<{$}  >{$}l<{$}  >{$}l<{$}  >{$}l<{$}  >{$}l<{$} }  
\toprule[1.5pt]
\hline
 & {\text{MSR}} &   {T_2} & {T_3} & {T_4} & {\text{DET}}  & {\text{LRF}}\\
\hline
\hline
\multicolumn{7}{l} {\textbf{E0}: Theoretical Value}\\
\hline
\STE{\tau}&0.8645&1338.3&10069 &8.35\text{E}4&48.322&73.678\\
$\Fdata {D}_{\text{T}}({\tau})$&-     &665.26 &93468 &1.30\text{E}7&1.3532&1.4210\\
$\Fdata {D}_{\text{1}}({\tau})$&4.4\text{E}\!-\!6&62.550&22197&4.25\text{E}6&0.3589&0.3646\\
$c_v$     &0.0024&0.0059&0.0148&0.0274&0.0124&0.0082\\
\toprule[1pt]

\multicolumn{7}{l} {\textbf{E1} [independent, 40]: No fluctuations around 18 MW}\\
\hline
$\mu_0$&1.0001&0.9909&0.9865&0.9775&0.9885&0.9932\\
$c_{\text{T0}}$  &-      &0.8832&0.9743&1.0453&0.8525&0.8461\\
\toprule[1pt]

\multicolumn{7}{l} {\textbf{E2} [independent, 40]: Small fluctuations around 18 MW}\\
\hline
$\mu_0$&0.9382&1.2091&1.6950&2.6860&0.6465&1.2196\\
$c_{\text{T0}}$  &-     &1.8532&2.2643&2.7016&2.1945&1.2051\\
\toprule[1pt]

\multicolumn{7}{l} {\textbf{E3} [independent, 40]: Small fluctuations around 300 MW}\\
\hline
$\mu_0$&0.9064&1.2765&2.0098&3.7752&0.4095&1.3753\\
$c_\text{{T0}}$  &-      &2.0341&2.7852&3.5392&3.2977&0.9713\\
\toprule[1pt]

\multicolumn{7}{l} {\textbf{S1} [{0240:0400}, {161}]: No fluctuations around 18 MW}\\
\hline
$\mu_0$&1.0029&0.9856&0.9763&0.9678&1.0010&0.9861\\
$c_0$  &0.7022&0.5407&0.6013&0.6008&0.7150&0.7519\\
\toprule[1pt]

\multicolumn{7}{l} {\textbf{S2} [{0401:0639}, {239}]: Transition stage from \textbf{S1} to \textbf{S3}}\\
\hline
$\mu_0$&0.9714&1.1045&1.3517&1.8213&0.8314&1.0966\\
$c_0$  &6.0177&8.9959&9.9893&10.982&7.5281&5.6884\\
\toprule[1pt]

\multicolumn{7}{l} {\textbf{S3} [{0640:0800}, {161}]: Small fluctuations around 18 MW}\\
\hline
$\mu_0$&0.9518&1.1801&1.6208&2.5243&0.7239&1.1669\\
$c_0$  &0.6892&1.0035&1.1688&1.2319&1.0925&0.6587\\
\toprule[1pt]

\multicolumn{7}{l} {\textbf{S4} [{0801:1039}, {239}]: A step signal (18 MW $\uparrow$ 300 MW) is included}\\
\hline
$\mu_0$&0.6170&8.4518&153.47&2712.4&-3.081&3.6629\\
$c_0$  &20.335&22.620&13.930&10.413&-12.92&10.796\\
\toprule[1pt]
\multicolumn{7}{l} {\textbf{S5} [{1040:1200}, {161}]: Small fluctuations around 300 MW}\\
\hline
$\mu_0$&0.9051&1.3204&2.3883&5.5954&0.3439&1.4169\\
$c_0$  &0.9242&1.4981&1.8841&2.2357&2.5537&0.6236\\
\toprule[1pt]
\multicolumn{7}{l} {\textbf{S6} [{1201:1377}, {177}]: A ramp signal (300 MW $\nearrow$ 359 MW) is included}\\
\hline
$\mu_0$&0.7921&3.4210&33.153&377.95&-0.789&2.1595\\
$c_0$  &32.664&66.503&46.417&33.682&-65.47&23.718\\
\toprule[1pt]
\multicolumn{7}{l} {\textbf{S7} [{1378:1600}, {223}]: Static voltage collapse}\\
\hline
$\mu_0$&0.3976&10.832&239.38&5831.3&-16.98&12.781\\
$c_0$  &91.867&100.67&70.117&51.575&-22.80&30.070\\
\toprule[1pt]
%

\hline
\toprule[1pt]
\end{tabularx}
\raggedright
 {*$\Equs{c_{v}}{\sqrt{\Fdata {D}_{\text{1}}({\tau})}/\Fdata {E}({\tau})}$ is the coefficient of variation;\\
\Equs{c_{\text{T0}}}{\Sig{\tau}/{\sqrt{\Fdata {D}_{\text{T}}}}}, \Equs{c_0}{\Sig{\tau}/\mu{(\tau)}/c_v}, and
\Equs{\mu_0}{{\Mu{\tau}}/\STE{\tau}}.

}
\end{minipage}
\end{table}

We summarize the performance of the LES system in SA:\\
a). The LES system is effective for stability evaluation. By comparing the deviation between the experimental value and the theoretical/empirical value for different system operating periods ($\textbf{E1}\!<\!\textbf{E2}\!<\!\textbf{E3}$), it come to a conclusion that \emph{the more steady the system is, the less the deviation becomes}. \\
b). When the system is under equilibrium operating situations without anomaly events occurring (\textbf{E1, E2, E3}), its LESs $\tau$ follow a Gaussian distribution,\footnote{Tested via jbtest Function in Matlab.} which in turn validates the theories in \eqref{eq:LESYYYY}.\\
c). The RMMs and LESs are able to be linked to some novel models, such as the \emph{ matrix-valued Brownian Movement}.  It is another topic and we will not expand here. \\
d). Different test functions  $\varphi({\lambda})$ have different characteristics and effectiveness; the signal may be sensitive to some test functions but not others. A test function is akin to a filter in some sense. Hence, we can balance the reliability and sensitivity for the detection in a special circumstance. Moreover, beyond benefiting anomaly detection (aiming to distinguish signals from noises), the test functions give us much more selections \emph{to handle the complicate situation} and have the potential \emph{to trace a specific anomaly} (aiming to distinguish the mixed signals).\\
e). Low $c_v$ means high precision and repeatability of the assay \cite{nutter1995improving}; from the aspect of ${c_v}$, MSR performs best. Especially, for a special purpose, e.g. the lowest  ${c_v}$  or the lowest bias, there  exists an optimal combination of Chebyshev Polynomials for the test function.\\

\subsection{Other LES Functions and the Visualization}
We can also utilize LESs to implement some other concrete functions, e.g., fault diagnosis.
By introducing the augmented matrix (\Equs{\Vector A}{[\Vector B ; \Vector C]}) which consists of the basic status data ($\Vector B$) and the factor data ($\Vector C$), our previous work \cite{he2015corr} conducts the correlation analysis with the LESs  of \Vector A and \Vector B. It is a feasible data-driven approach to find out the causing factor of the anomaly with no knowledge in advance.

The LES is a statistical indicator; its value depends on a large number of sample data points ${\Rdata x_{i,j}}$   in the form of the entries of the matrix \VRX{}. It means that LES $\tau{}$ is universal with the least assumptions and robust against individual bad data. Then, a case is designed to validate the robustness. Furthermore, the case also gives the visualization of the system from LES indicator perspective; meanwhile,  a comparison with perspective of low-dimensional indicator, e.g., raw data $\hat {\mathbf{V}}$,  is made.

We keep the assumed scenario in Table \ref{Tab: Event Series}, and take \VLES{LRF} as the test function.
 We obtain a local \Equs{\mu_0}{\VLES{LRF}/\STE{\VLES{LRF}}} for each region in a similar way to that of the whole system (118 nodes); thus, the \Cur {\mu_0}{t} curves are plotted as Fig. \ref{fig:Ples}. The division depends on the specific network structure; however, a potential problem lies here: how small the partition  (Fig. \ref{fig:IEEE118network}) can be that keeps the theories still work.
To find the answer we should study how to turn big data into tiny data \cite {feldman2013turning}; it is another topic and we do not expand here.

\begin{figure}[htbp]
\centering
\begin{overpic}[scale=0.36
]{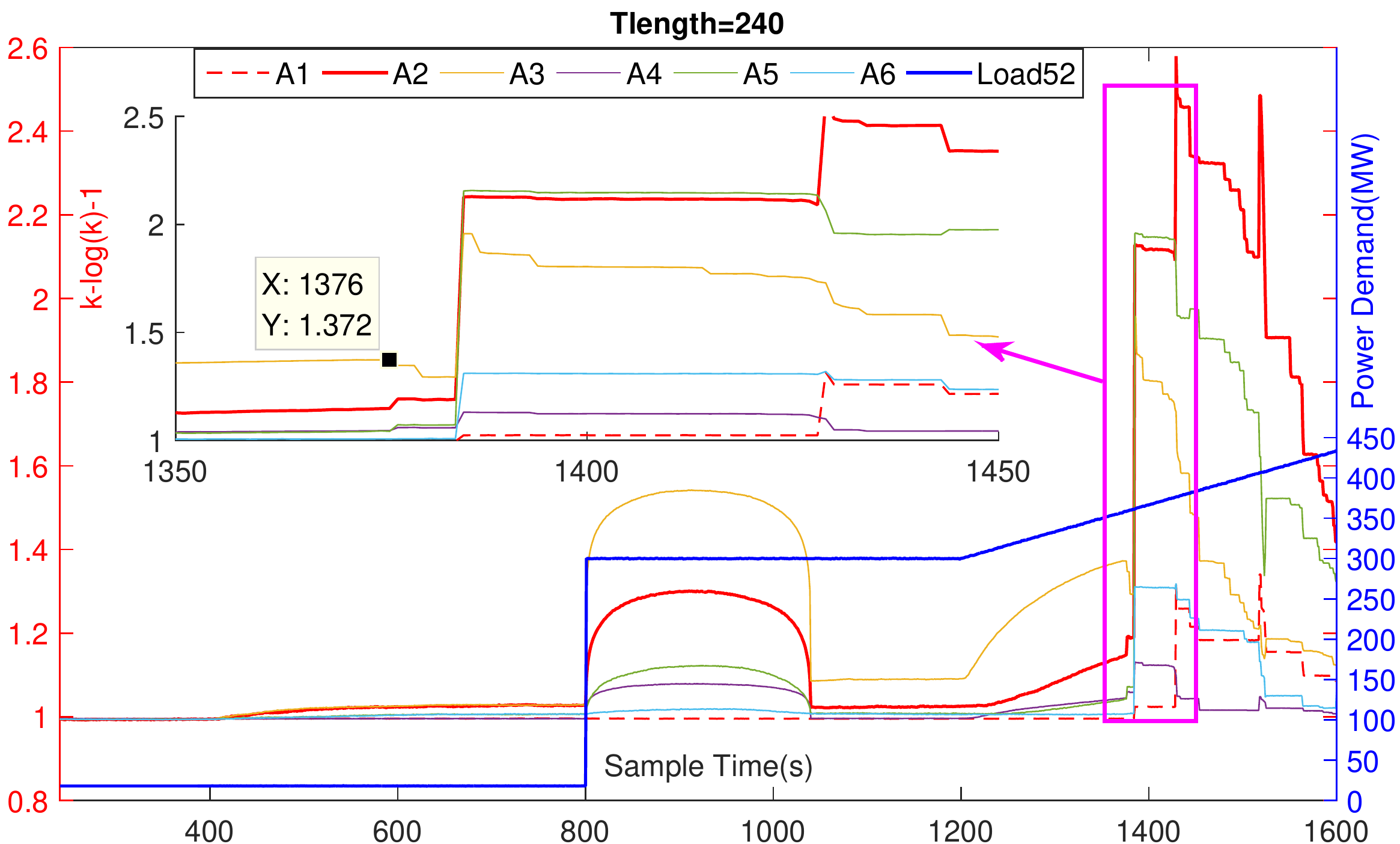}
\end{overpic}
\caption{\Cur {\mu_0}{t} Curve for Single Partitioning: A1--A6}
\label{fig:Ples}
\end{figure}

 With an interpolation method \cite{weber2000voltage}, a 3D~power map is plotted as the illustration of $\mu_0$; then the animation is produced.
Fig. \ref{fig:FullDataMSR} and Fig. \ref{fig:DataWithoutA3MSR} depict some key frames of the animation about LES indicator $\mu_0$, whereas Fig. \ref{fig:FullDataV} and Fig. \ref{fig:DataWithoutA3V} about raw data  $\hat {\mathbf{V}}$. From Fig. \ref{fig:Ples} and the animation, we obtain:
\\
{}
a) In Fig \ref{fig:FullDataMSR}, at \Et{801}{s}, {$\mu_0$} of A3 area changes dramatically and keeps for  $T$  sampling points.  The duration $T$ is observed in the animation; it is also reflected by Fig. \ref{fig:Ples}}.  Therefore, we deduce that some event occurs in A3; even we can go further that the event is a regional one and influential to A1, A2, A4, A5, and has little impact on A6.  These conjectures, in a reasonable way, coincide with the common sense that there is a sudden load change in A3 at \Et{801}{s} as we assumed in Table  \ref{Tab: Event Series}.\\
b) With sustainable growth of \VPbus{52}, the whole system becomes more and more vulnerable. The vulnerability, before the system has a breakdown due to voltage collapse,  can be estimated in advance via the animation of $\mu_0$.\\
c) Subfigures of (c), (d), (e) are almost the same for Fig. \ref{fig:DataWithoutA3V}, but not for Fig. \ref{fig:DataWithoutA3MSR}; the animation demonstrates this phenomenon  more clearly. That means, even with the \emph{most related data} (data of A3) losses which will lead to an insufficient  SA using  $\hat {\mathbf{V}}$,   the \emph{proper judgements} can still be achieved by $\mu{}_0$.\\
d) In this way, the interconnected grid is decoupled naturally. We can conduct regional SA only with its own datasets; these datasets are relatively independent. 

\begin{figure*}[htbp]
\centering
\subfloat[\EtS{800}{s}]{
\begin{overpic}[width=0.19\textwidth]{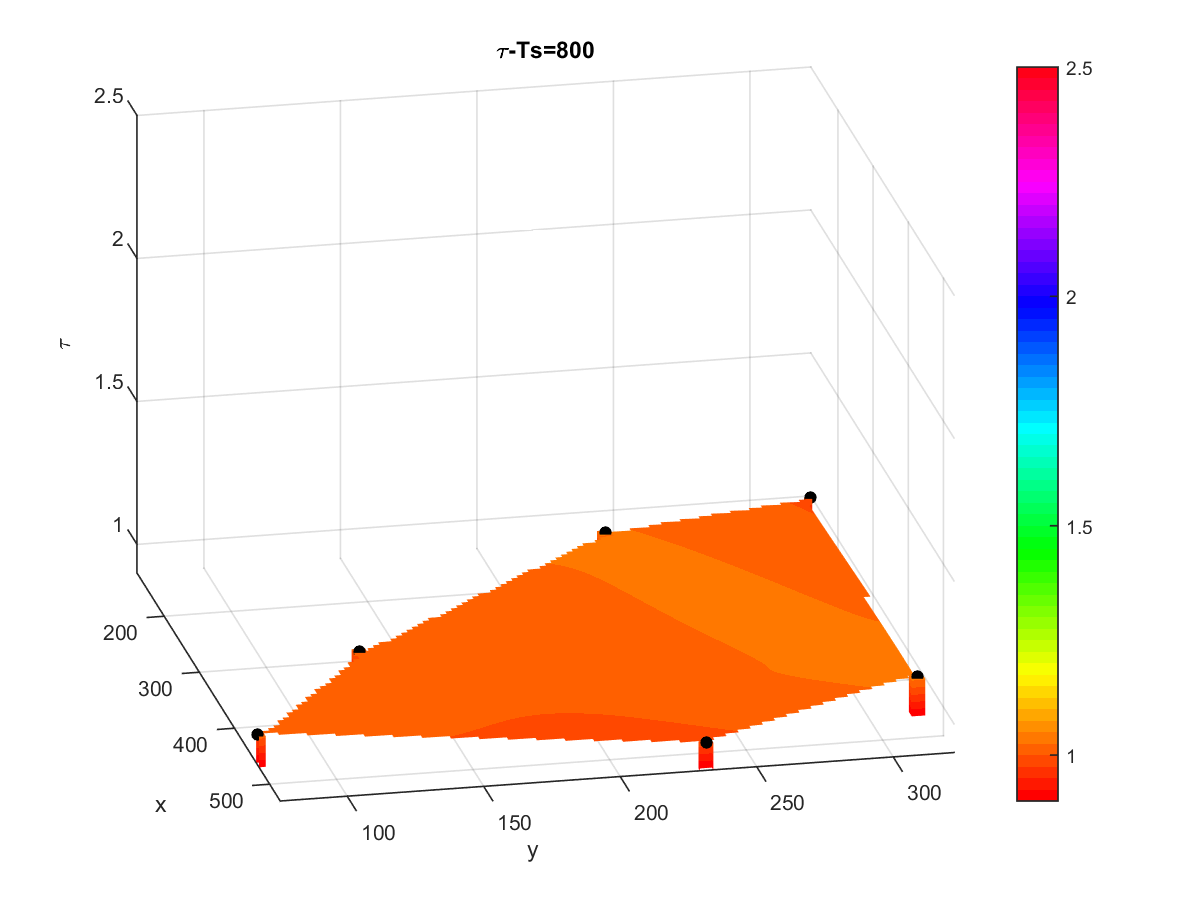}
    \setlength {\fboxsep}{1pt}
   \put(65,38) {\fbox{\tiny \color{blue}{\textbf A$1$}}}   
   \put(45,34) {\fbox{\tiny \color{blue}{\textbf A$2$}}}   
   \put(80,20) {\fbox{\tiny \color{blue}{\textbf A$3$}}}   
   \put(65,4) {\fbox{\tiny \color{blue}{\textbf A$4$}}}   
   \put(22,24) {\fbox{\tiny \color{blue}{\textbf A$5$}}}   
   \put(12,16) {\fbox{\tiny \color{blue}{\textbf A$6$}}}   
\end{overpic}
}
\subfloat[\EtS{801}{s}]{
\includegraphics[width=0.19\textwidth]{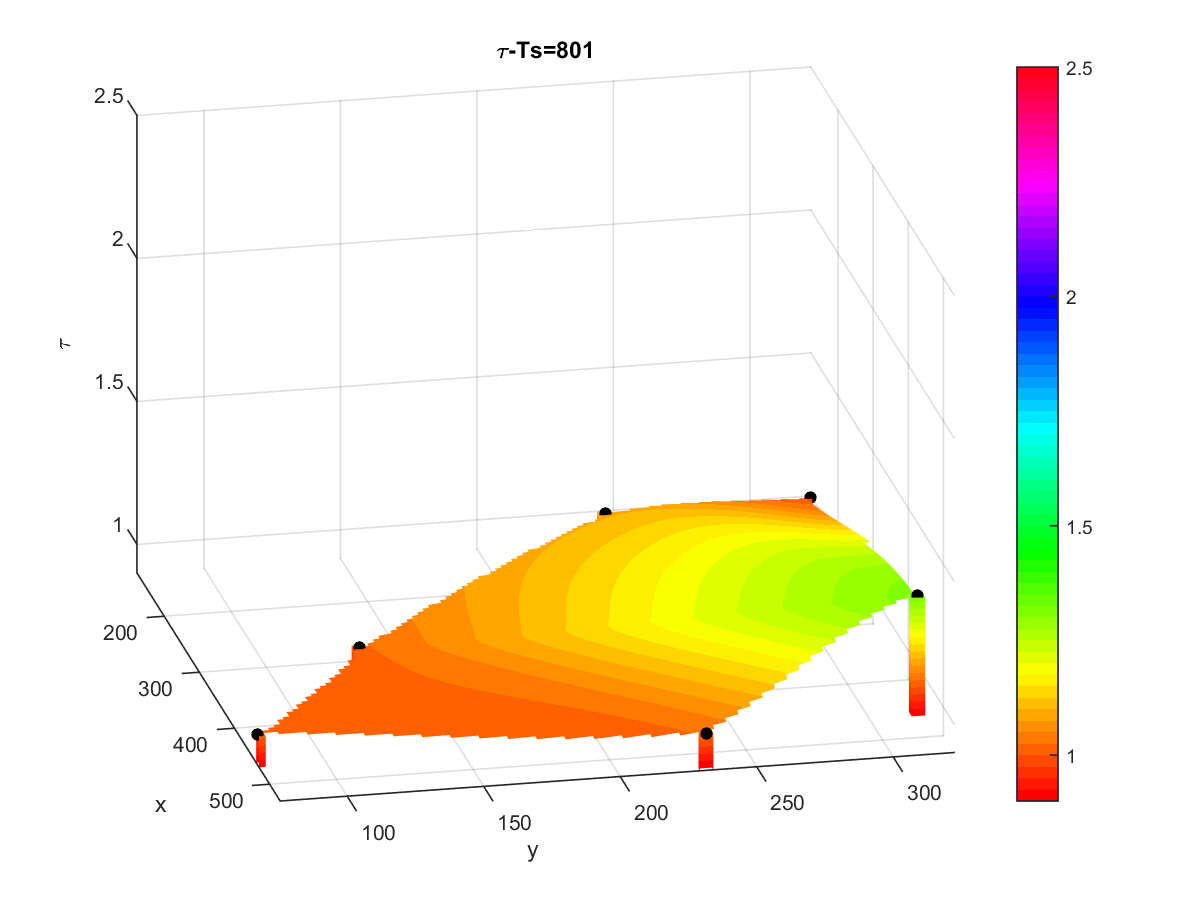}
}
\subfloat[\EtS{920}{s}]{
\includegraphics[width=0.19\textwidth]{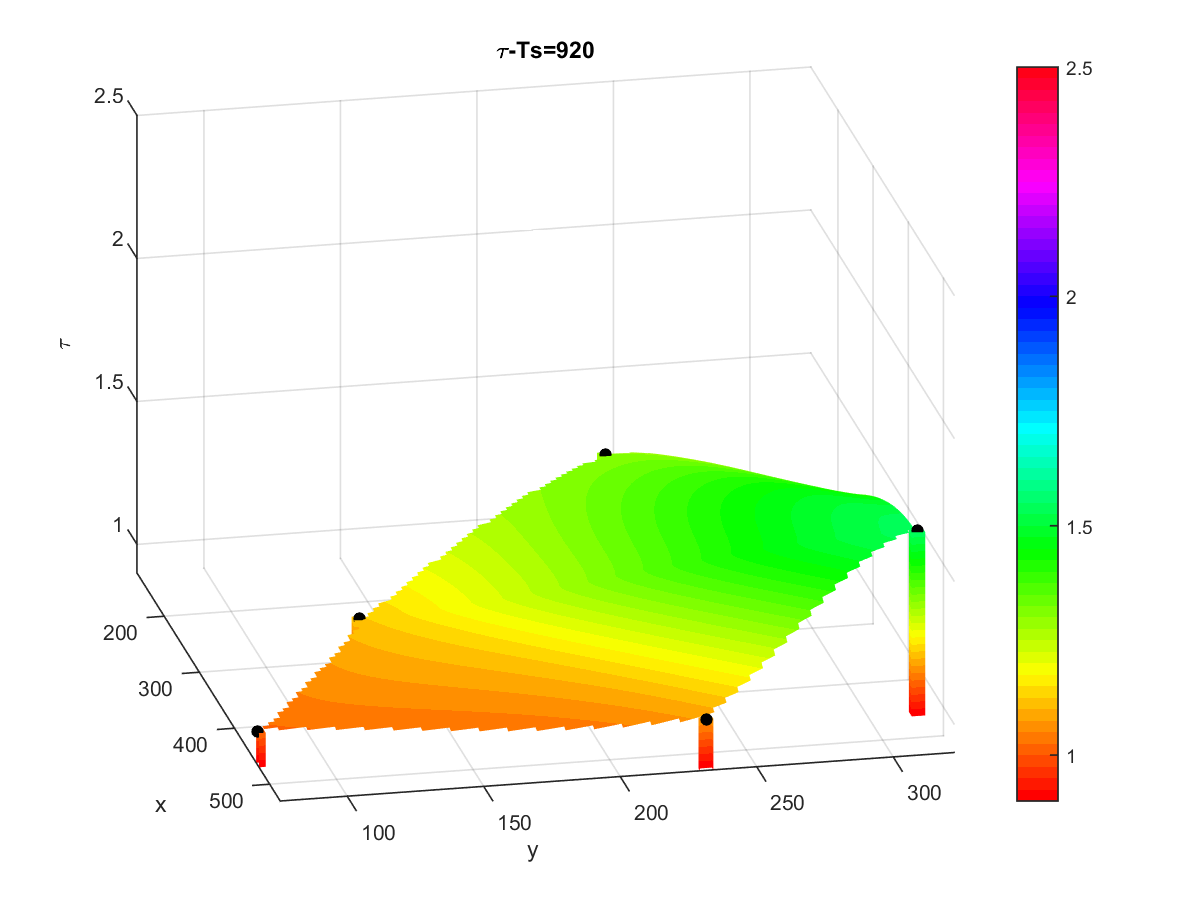}
}
\subfloat[\EtS{1360}{s}]{
\includegraphics[width=0.19\textwidth]{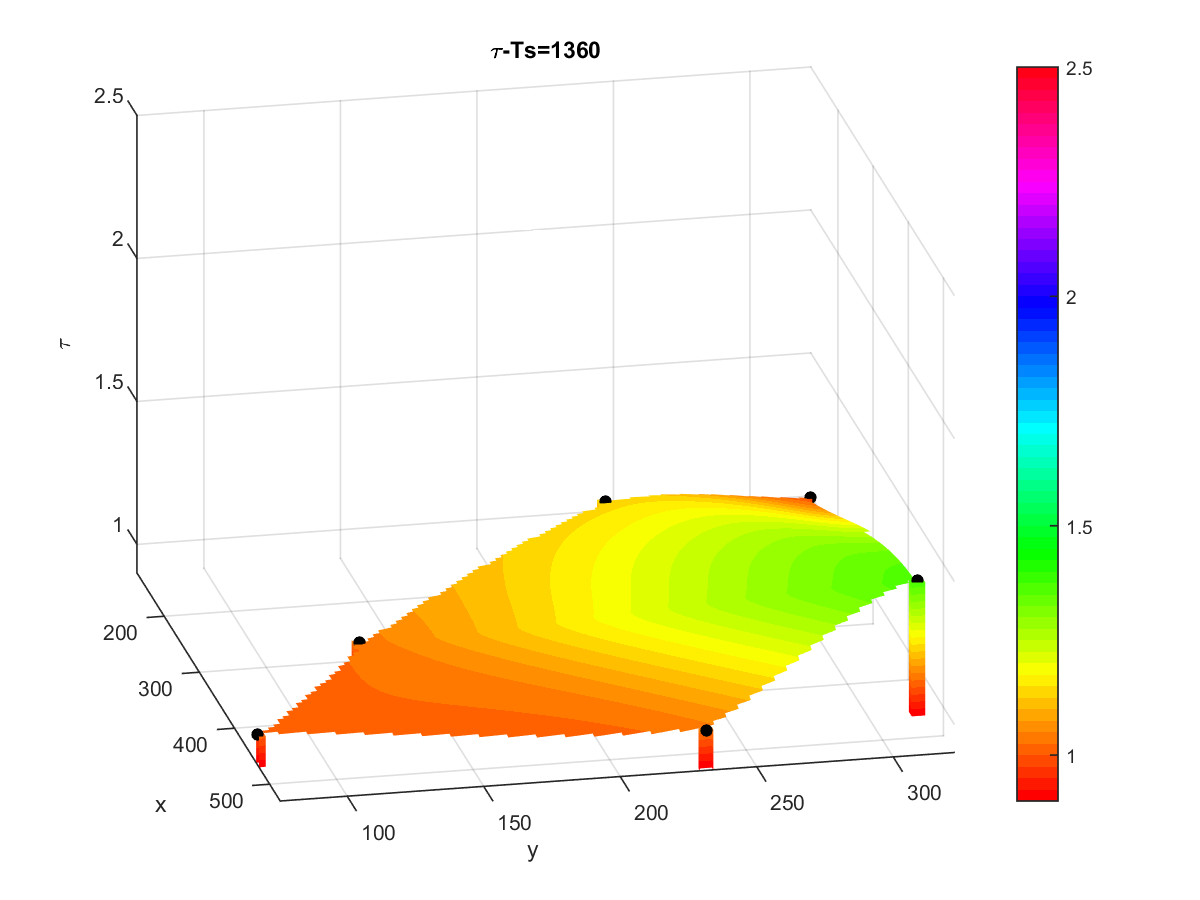}
}
\subfloat[\EtS{1450}{s}]{
\includegraphics[width=0.19\textwidth]{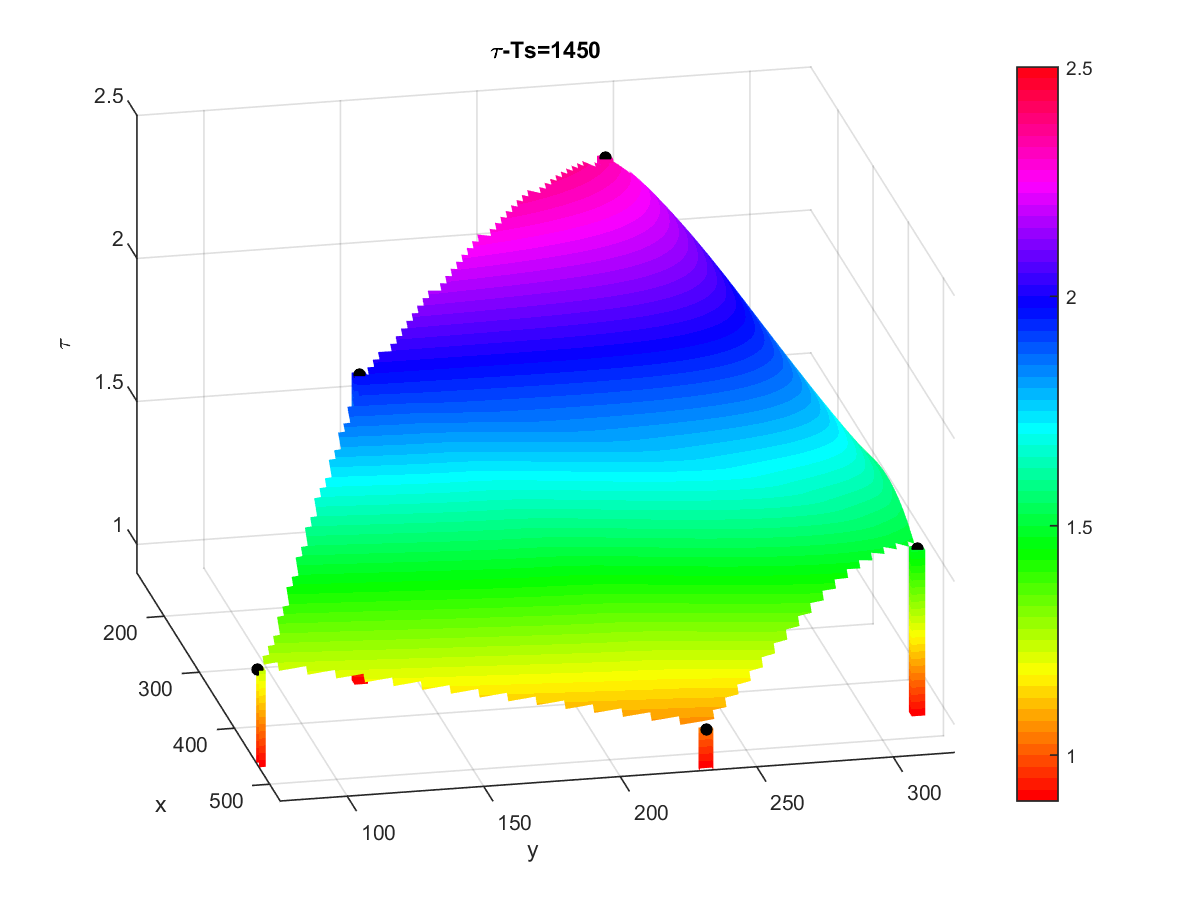}
}
\caption{Visualization of the high-dimensional indictor $\mu_0$ with Full Data Sets}
\label{fig:FullDataMSR}

\centering
\subfloat[\EtS{800}{s}]{
\includegraphics[width=0.19\textwidth]{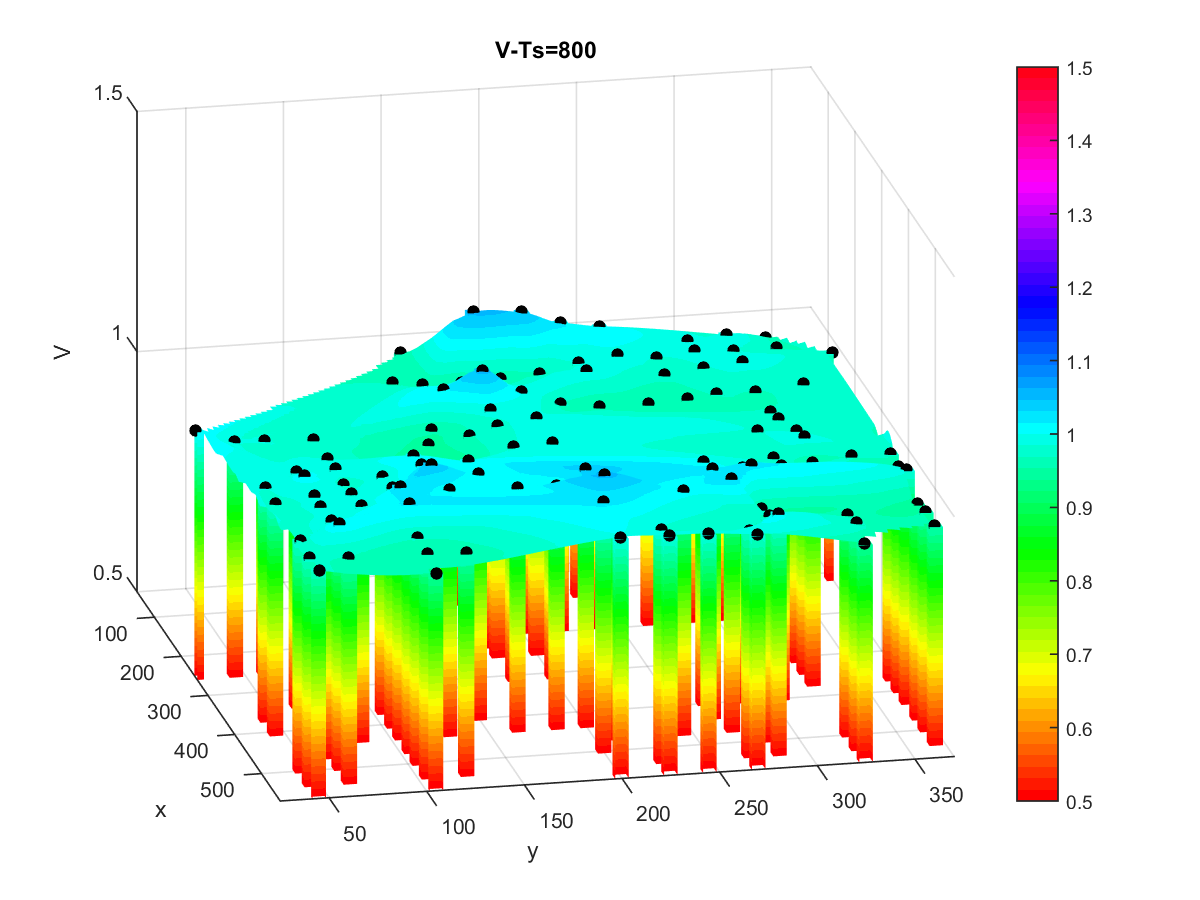}
}
\subfloat[\EtS{801}{s}]{
\includegraphics[width=0.19\textwidth]{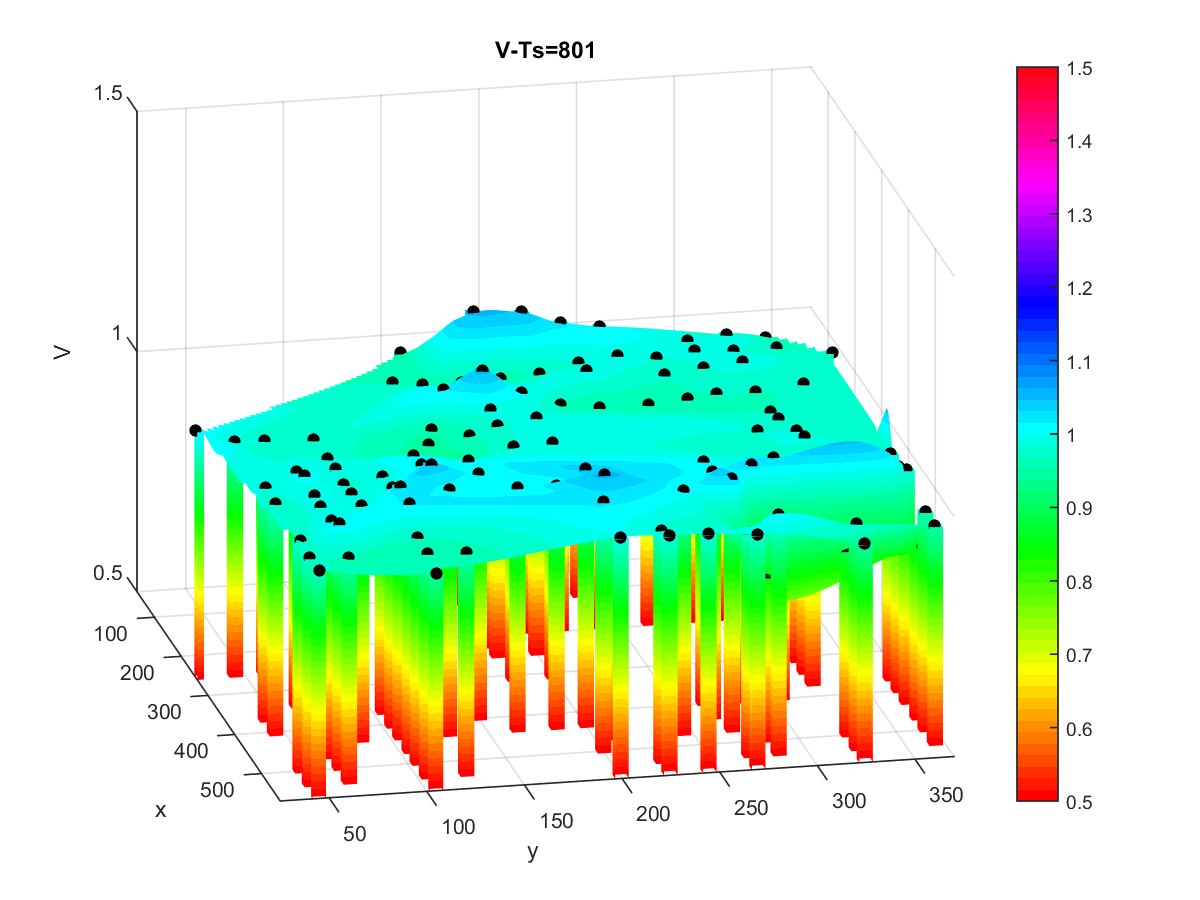}
}
\subfloat[\EtS{920}{s}]{
\includegraphics[width=0.19\textwidth]{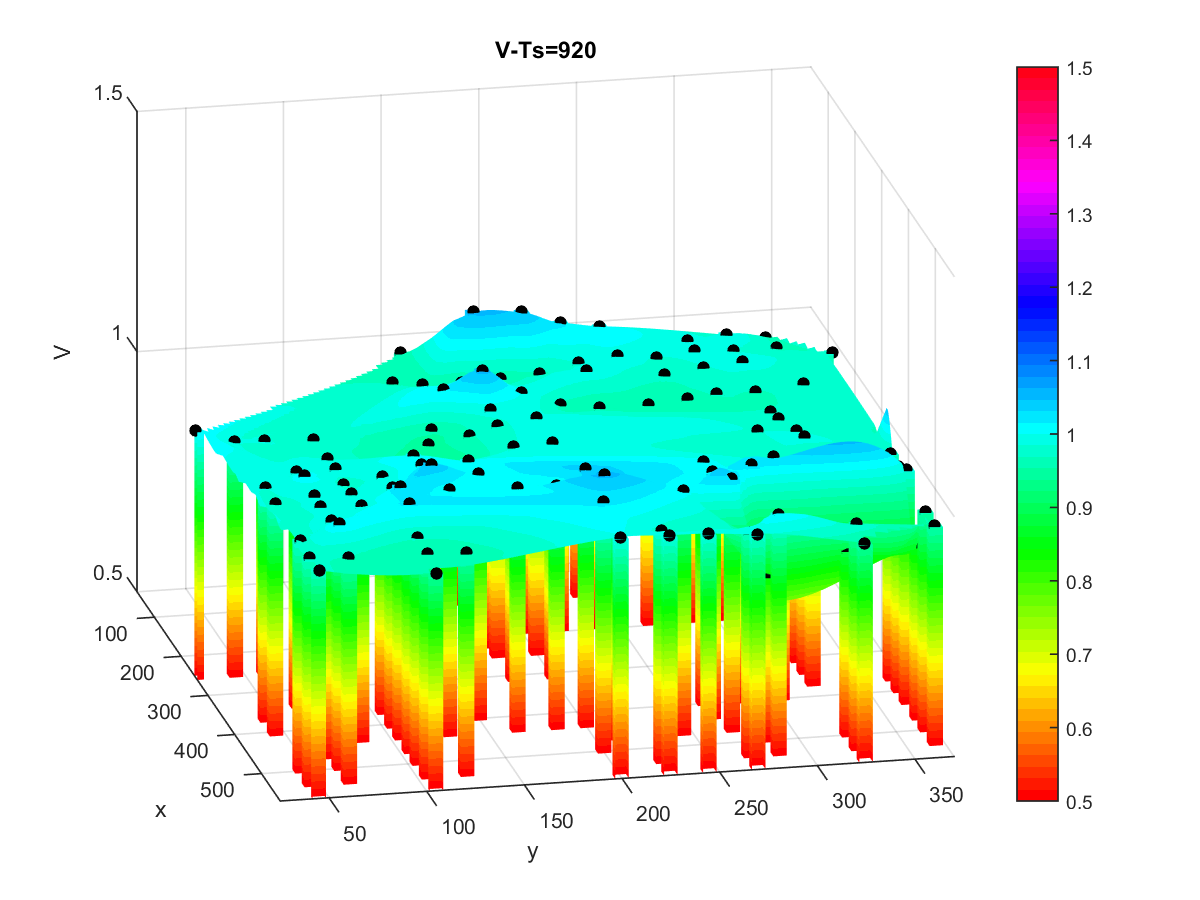}
}
\subfloat[\EtS{1360}{s}]{
\includegraphics[width=0.19\textwidth]{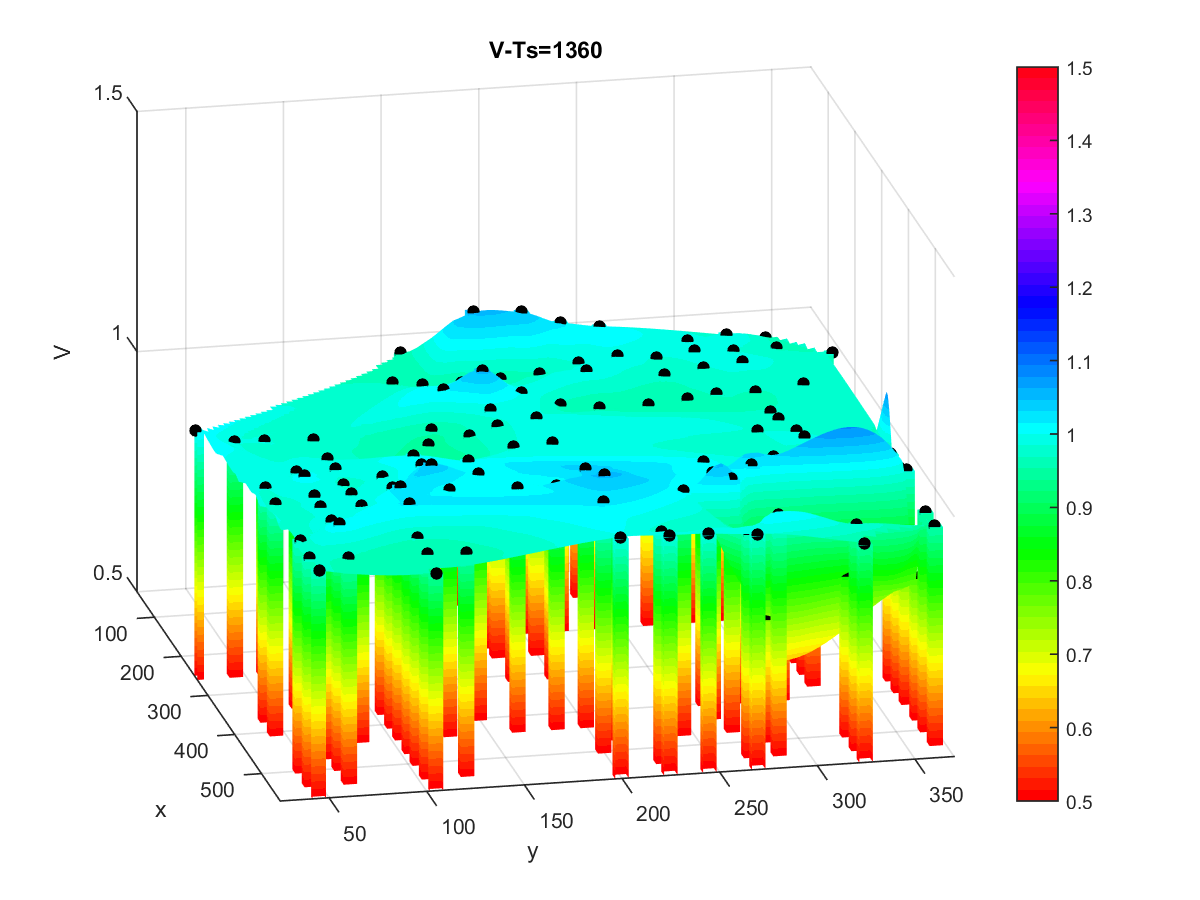}
}
\subfloat[\EtS{1450}{s}]{
\includegraphics[width=0.19\textwidth]{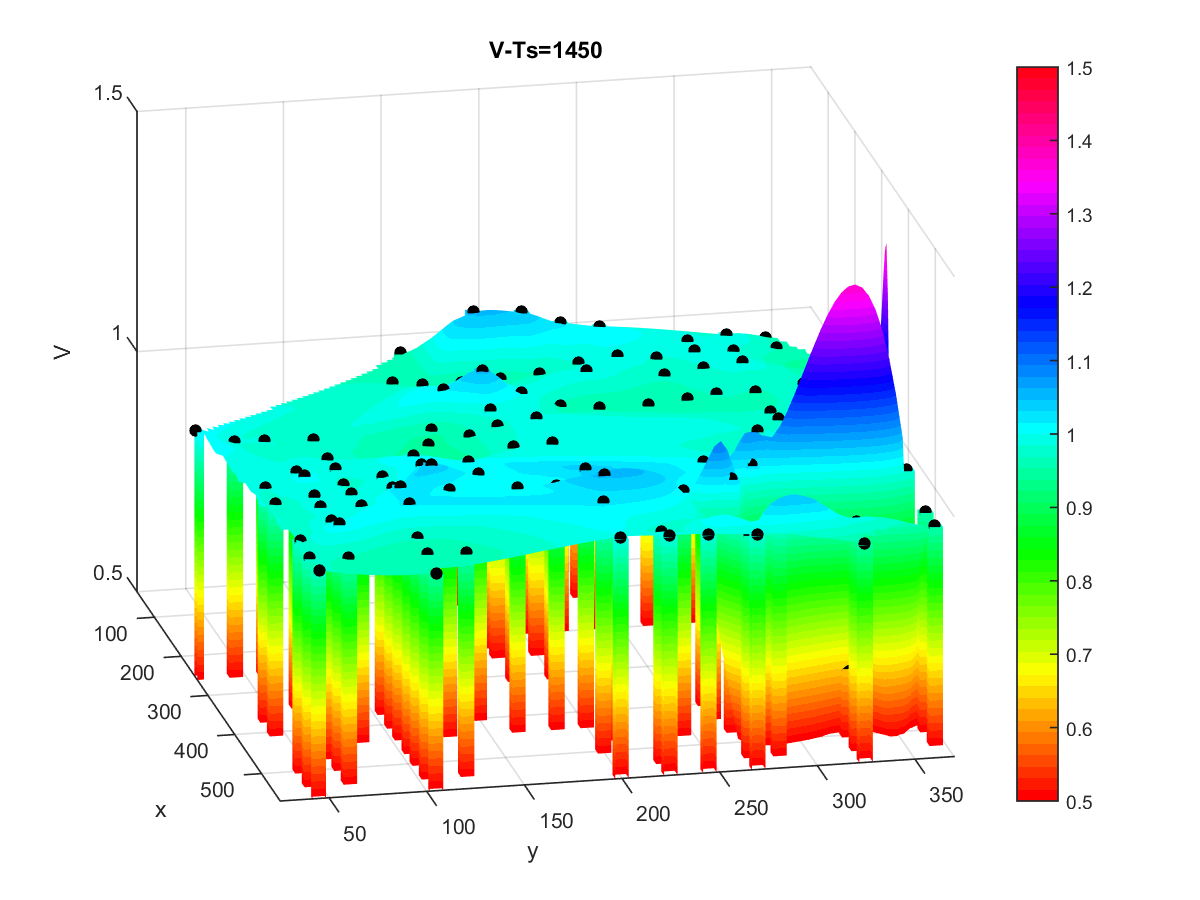}
}
\caption{Visualization of the Voltage $\hat {\mathbf{V}}$ with Full Data Sets}
\label{fig:FullDataV}

\centering
\subfloat[\EtS{800}{s}]{
\begin{overpic}[width=0.19\textwidth]{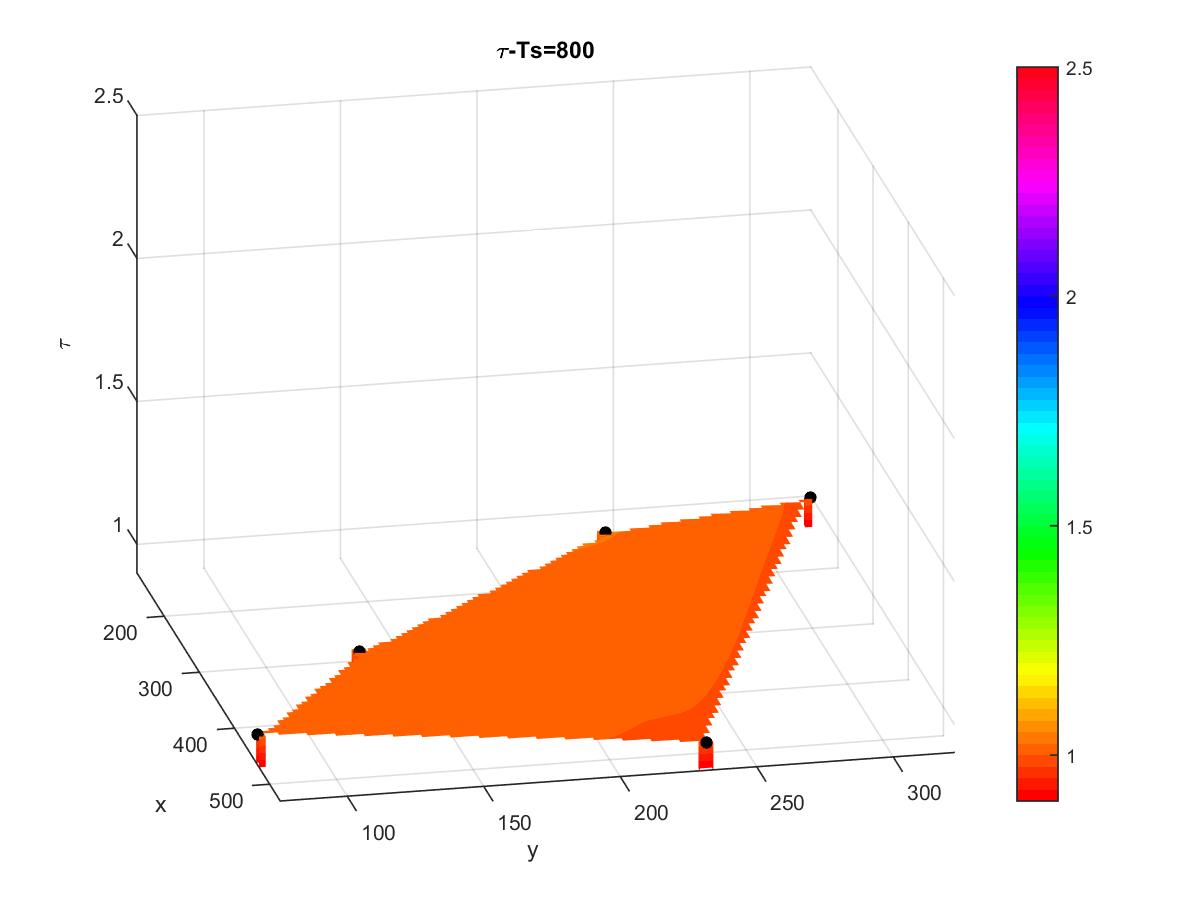}
    \setlength {\fboxsep}{1pt}
   \put(65,38) {\fbox{\tiny \color{blue}{\textbf A$1$}}}   
   \put(45,34) {\fbox{\tiny \color{blue}{\textbf A$2$}}}   
   \put(65,4) {\fbox{\tiny \color{blue}{\textbf A$4$}}}   
   \put(22,24) {\fbox{\tiny \color{blue}{\textbf A$5$}}}   
   \put(12,16) {\fbox{\tiny \color{blue}{\textbf A$6$}}}   
\end{overpic}
}
\subfloat[\EtS{801}{s}]{
\includegraphics[width=0.19\textwidth]{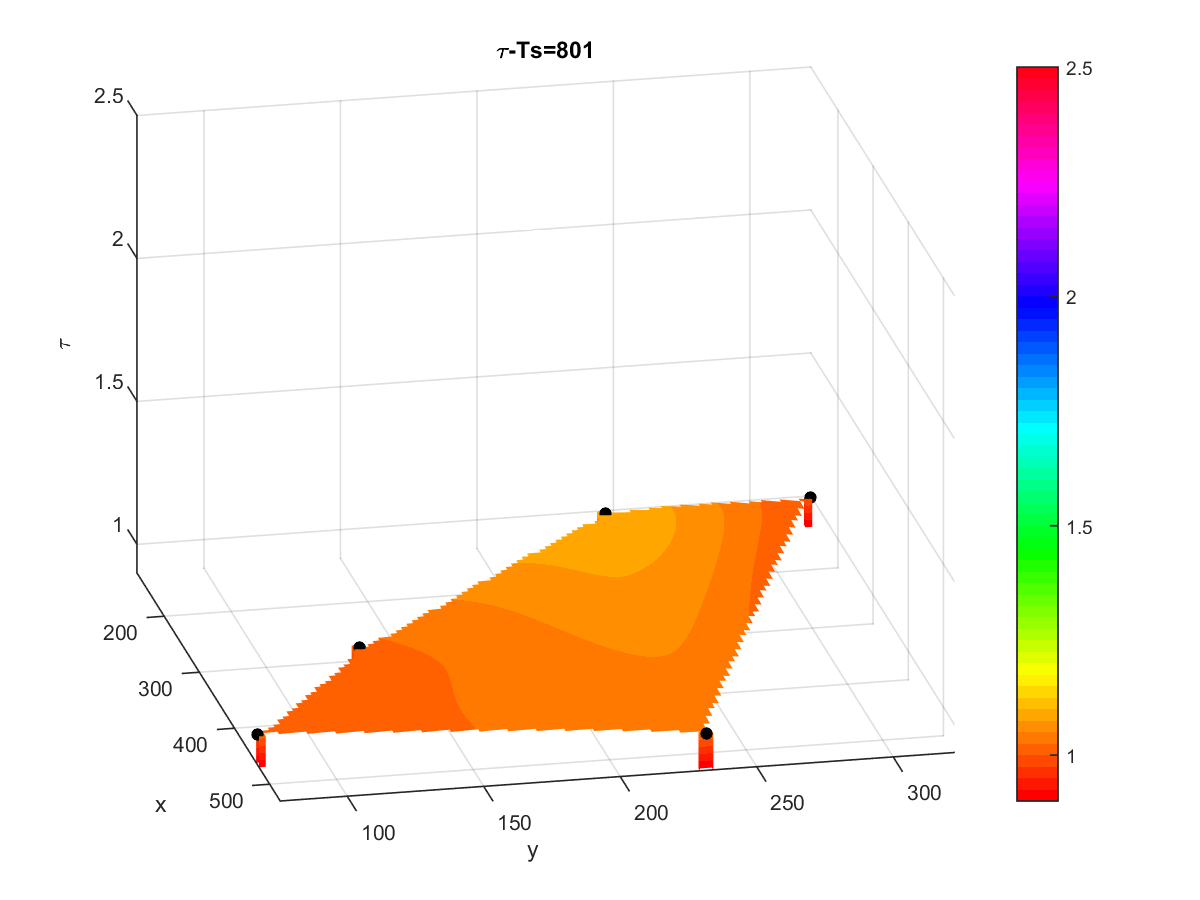}
}
\subfloat[\EtS{920}{s}]{
\includegraphics[width=0.19\textwidth]{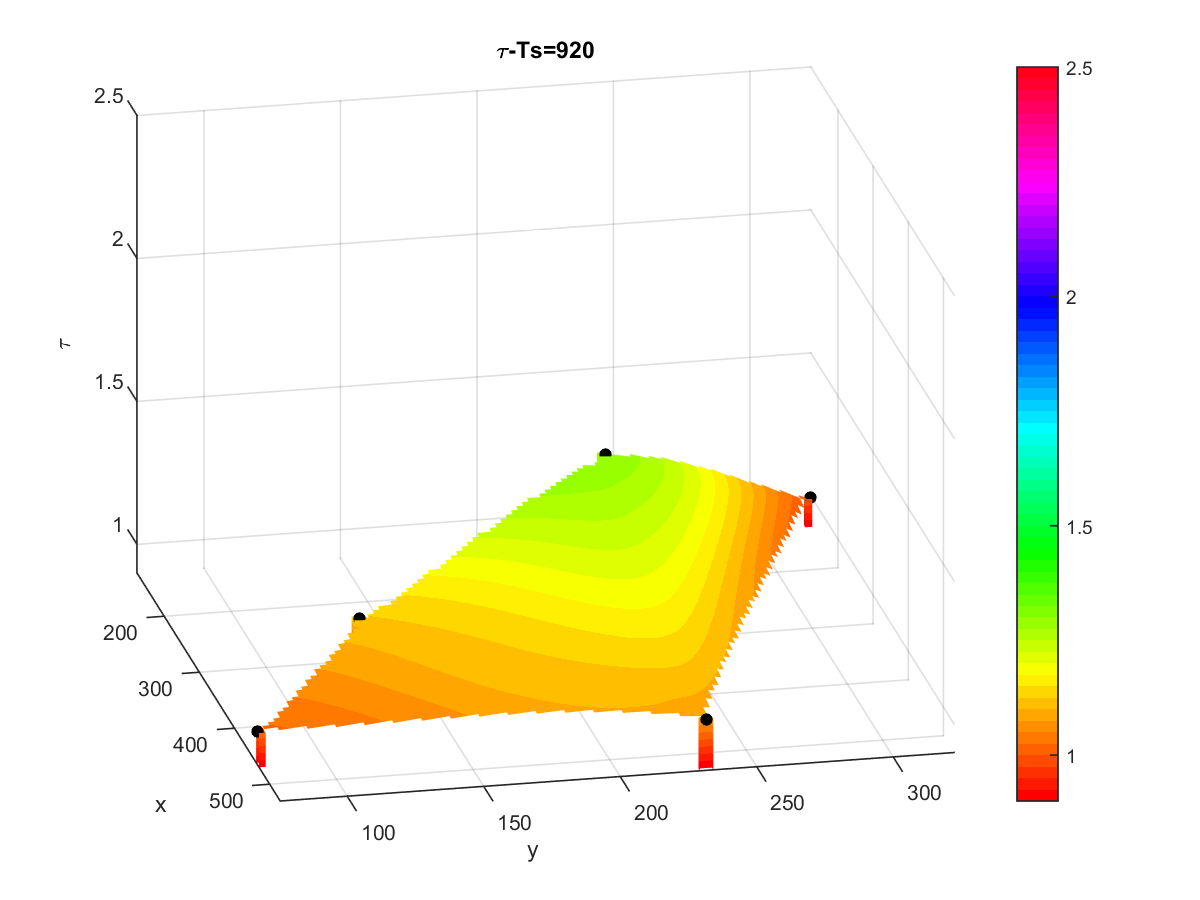}
}
\subfloat[\EtS{1360}{s}]{
\includegraphics[width=0.19\textwidth]{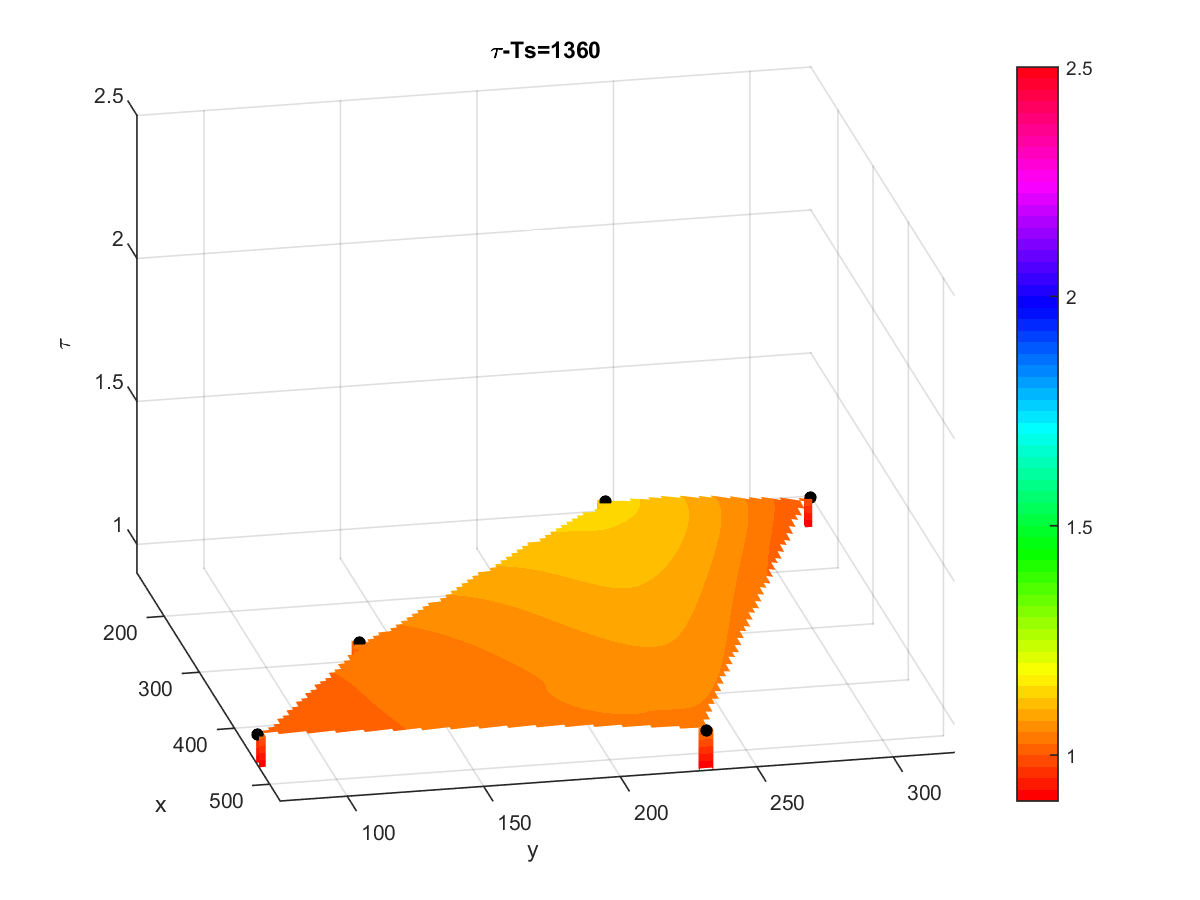}
}
\subfloat[\EtS{1450}{s}]{
\includegraphics[width=0.19\textwidth]{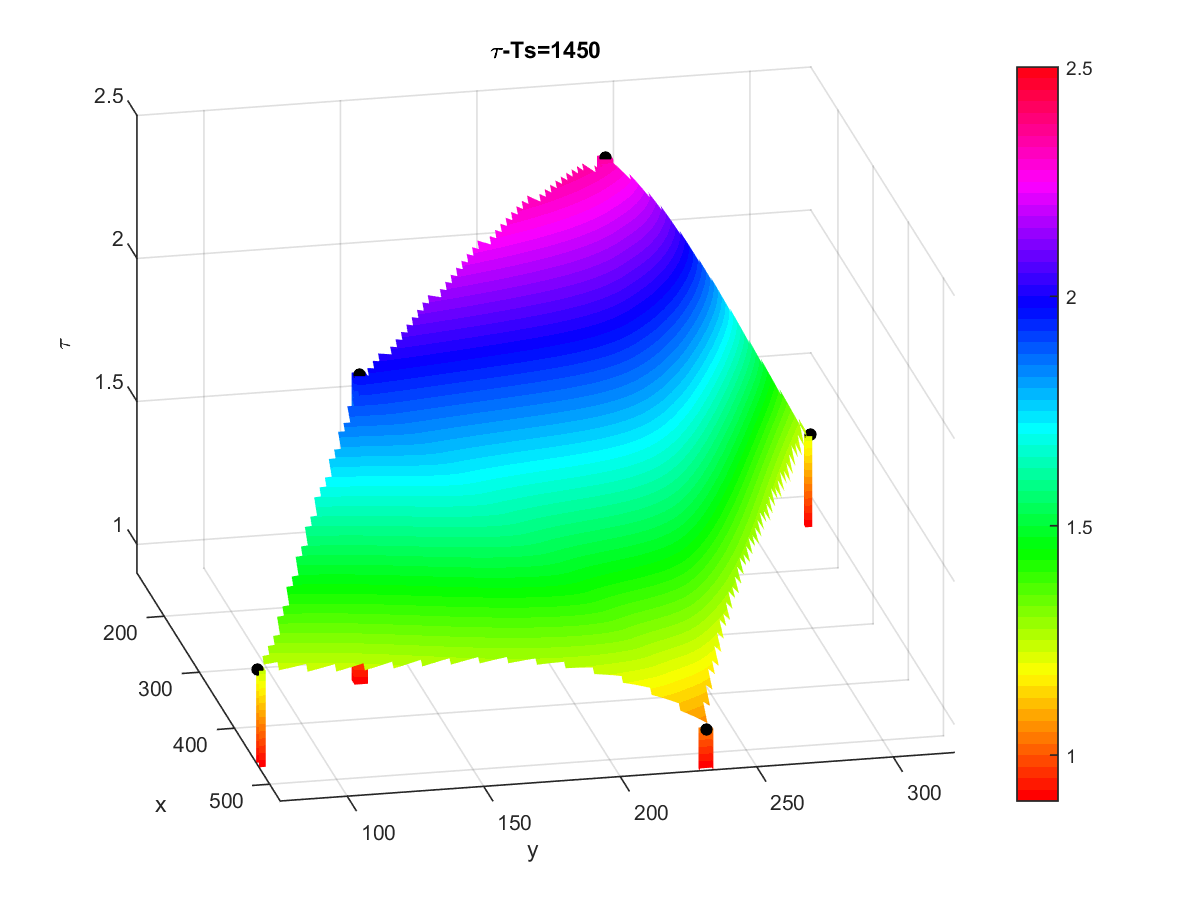}
}
\caption{Visualization of the high-dimensional indictor $\mu_0$ without Data Set of A3}
\label{fig:DataWithoutA3MSR}
\subfloat[\EtS{800}{s}]{
\includegraphics[width=0.19\textwidth]{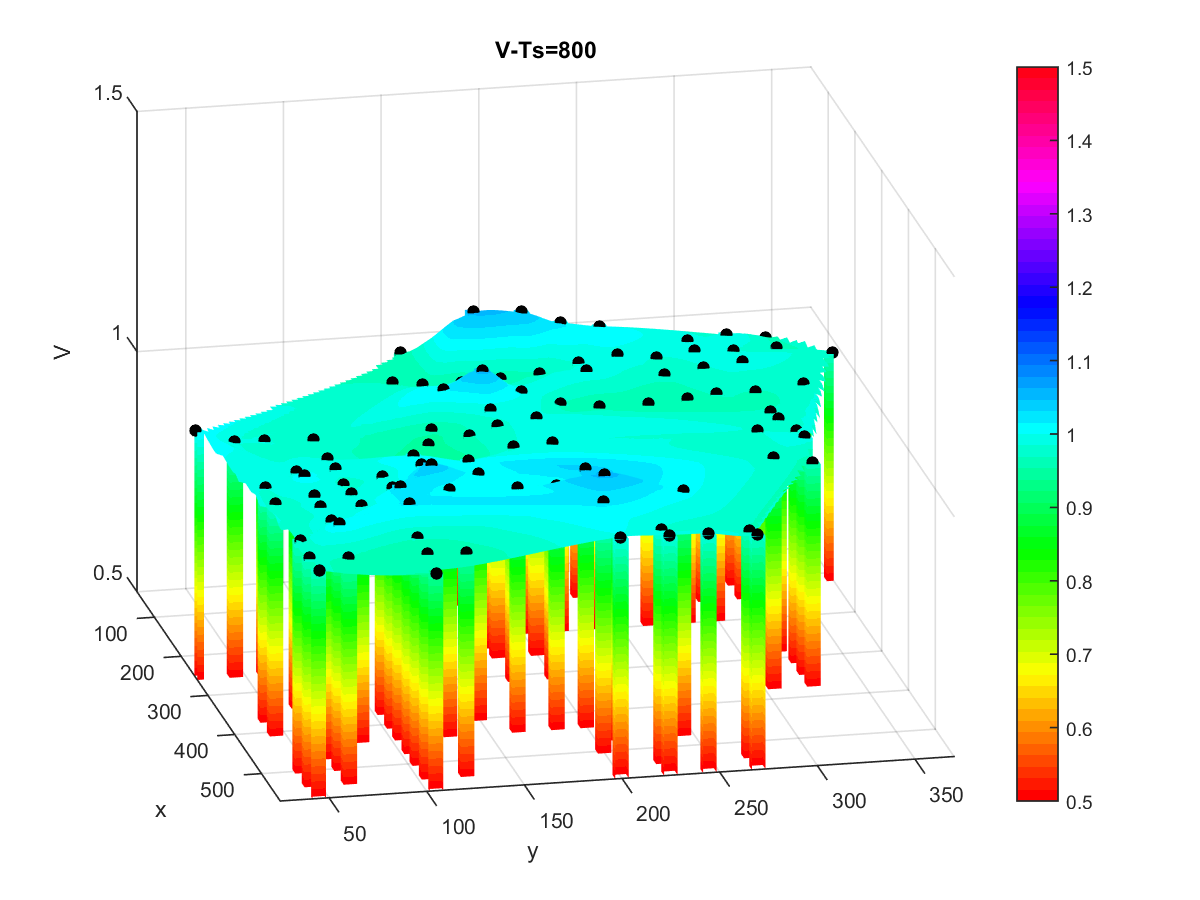}
}
\subfloat[\EtS{801}{s}]{
\includegraphics[width=0.19\textwidth]{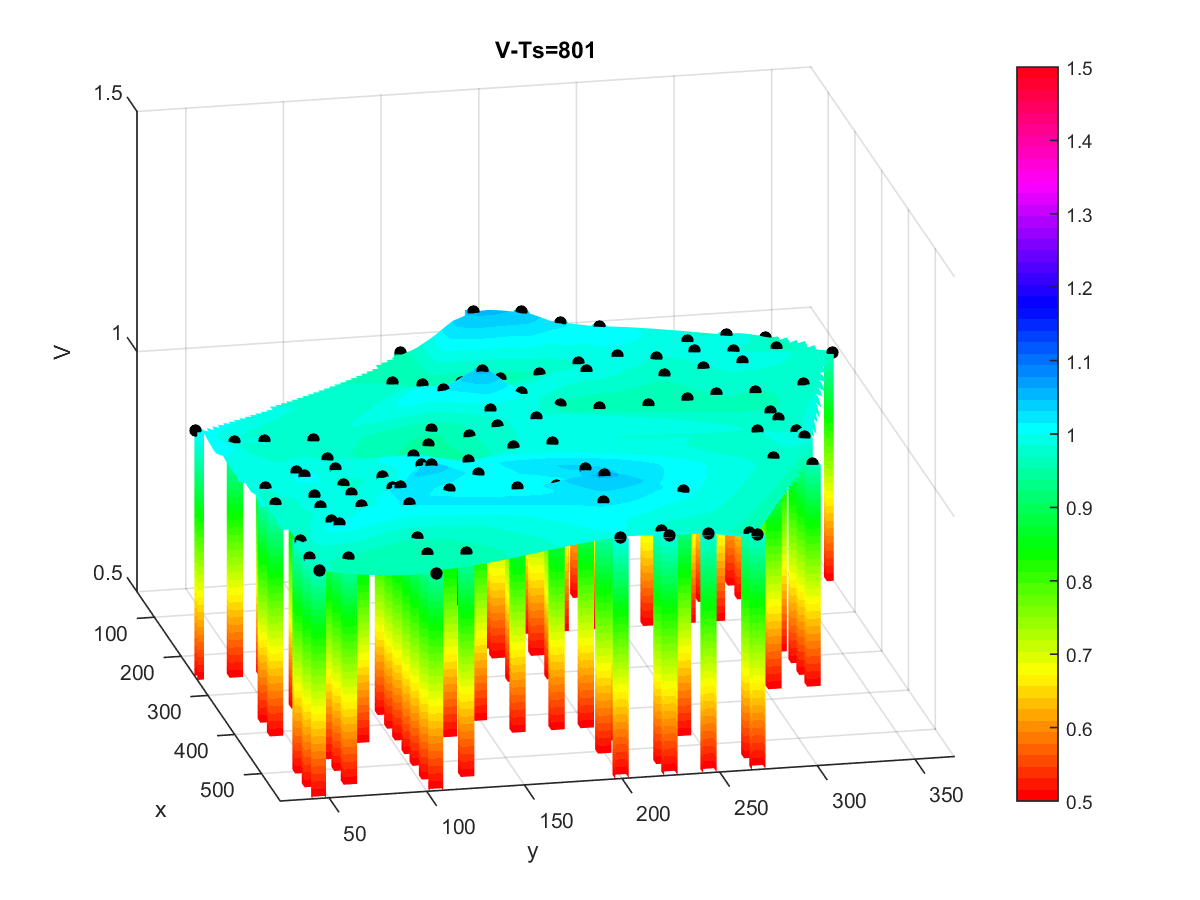}
}
\subfloat[\EtS{920}{s}]{
\includegraphics[width=0.19\textwidth]{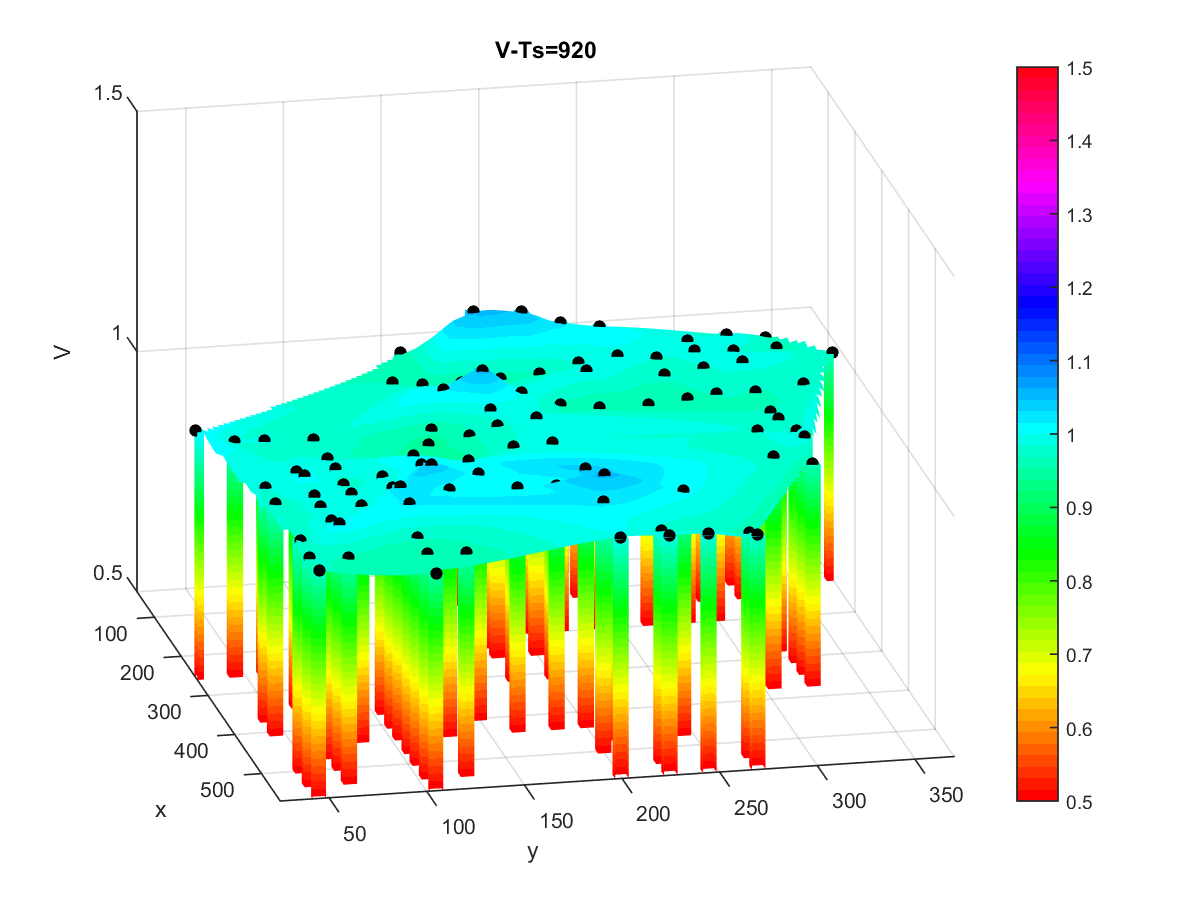}
}
\subfloat[\EtS{1360}{s}]{
\includegraphics[width=0.19\textwidth]{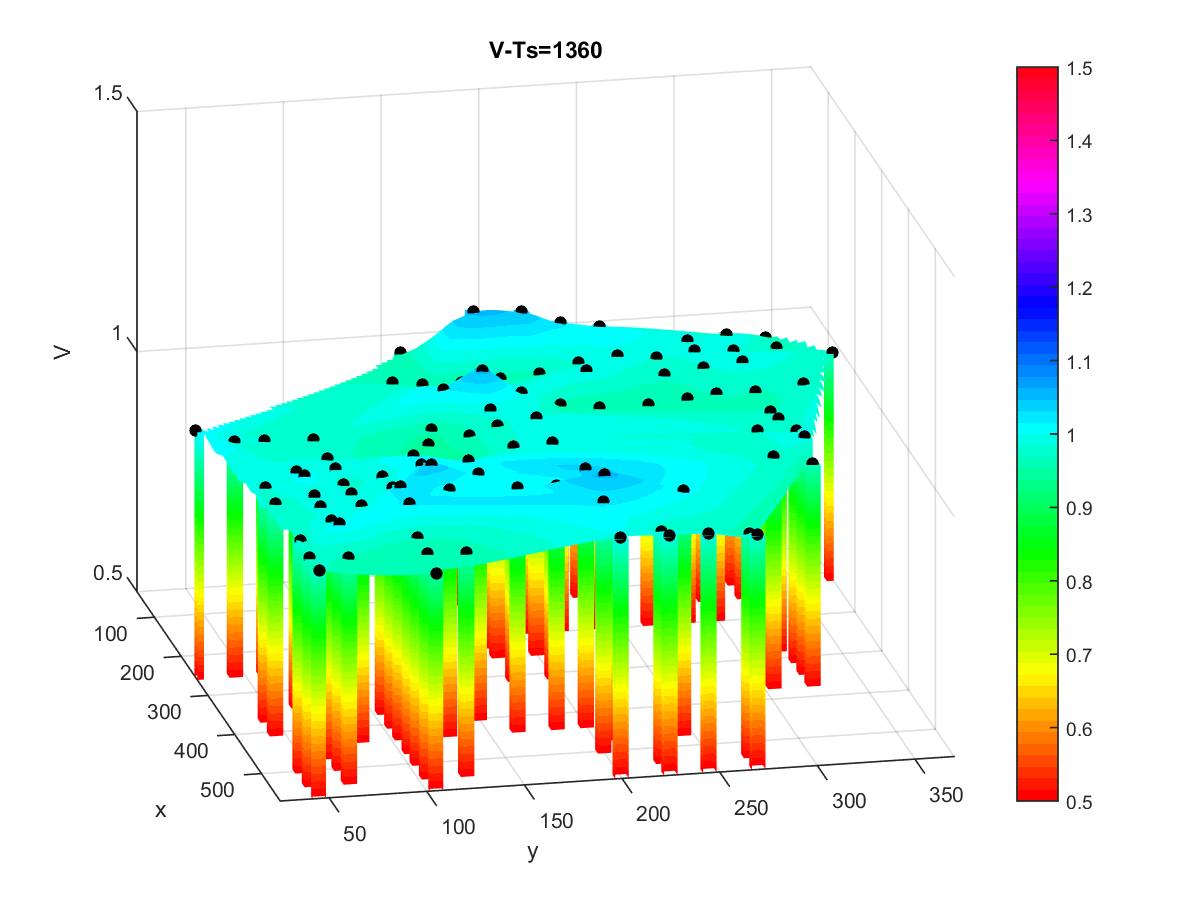}
}
\subfloat[\EtS{1450}{s}]{
\includegraphics[width=0.19\textwidth]{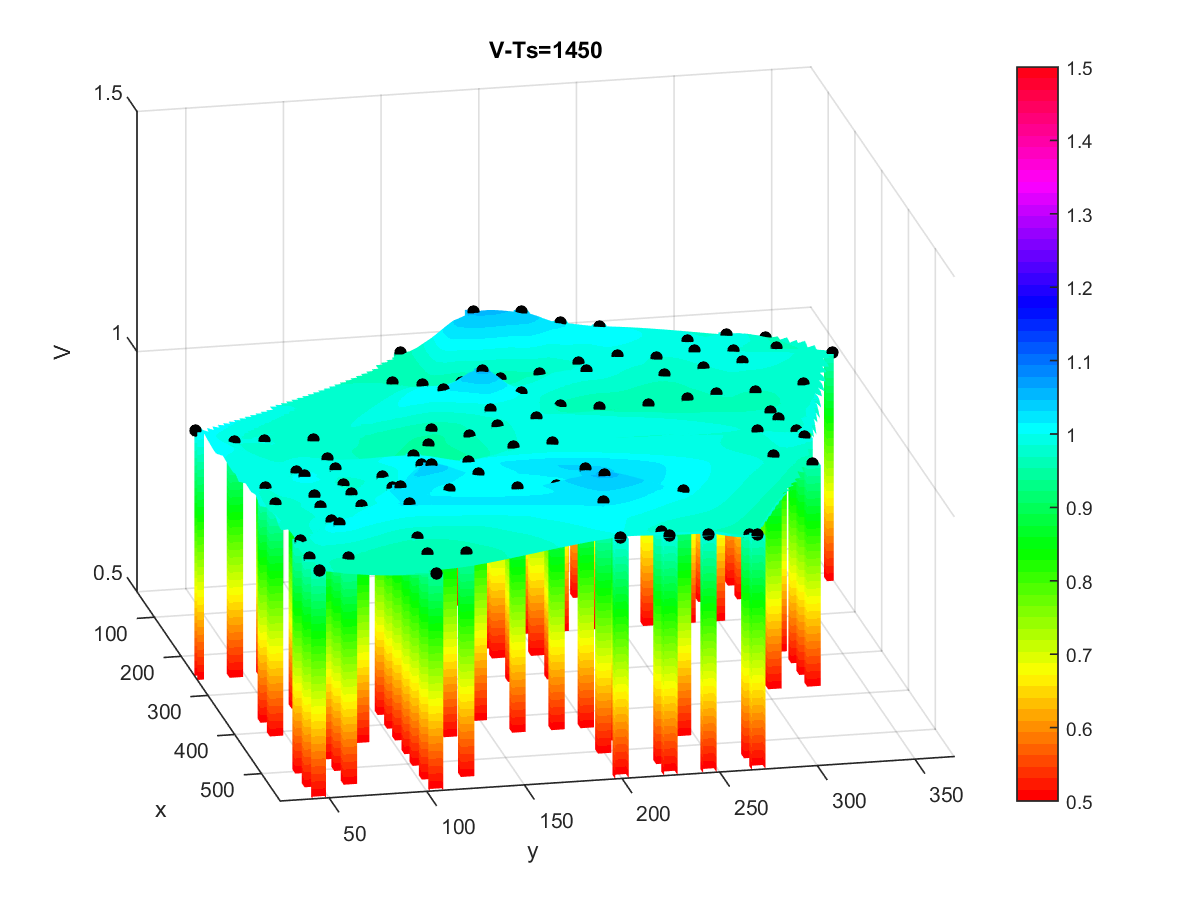}
}
\caption{Visualization of the Voltage $\hat {\mathbf{V}}$ without Data Set of A3}
\label{fig:DataWithoutA3V}
\end{figure*}

\section{Conclusion}
\label{section: concl}
\normalsize{}

This paper, motivated for the future's electrical grid, studies the situation awareness (SA) based on linear eigenvalue statistics (LESs).
Three ingredients are essential for the data-driven SA: 1)  data modeling---building the random matrix model (RMM) for a certain physical problem and employing the RMM to connect the sampling data with statistical theories/tools; 2) data analytics---conducting high-dimensional analysis to obtain the statistical indicators LESs; 3)  interpretation---interpreting and visualizing the indicators to gain insight into the system.

For engineering applications, based on  the experimental LESs which are fully derived from the sampling data, various SA functions are studied; for example, by comparing the LESs with their theoretical prediction, the anomaly detection  is conducted.  In addition, the 3D power-map animation is developed for the visualization.

Future research directions include: (1) Model validation with different implementations of the grid, ranging from statistic, dynamic and real-world systems; (2) Data fusion with a number of random data matrices, using mathematical tools such as free probability; (3) The use of Gaussian random matrices in replacement for general data matrices that are obtained from the electrical grid. The universality principle of random matrix theory says that this replacement causes negligible errors. (4) The convergence rate of the central limit theorem, using the Berry-Essen type of inequalities~\cite{Qiu2016BigDataLRM}, is used to study how fast the number of the data nodes (such as PMUs) converges to the limit.


\appendices

\section{}

\begin{figure}[H]
\centering
\begin{overpic}[scale=0.58
]{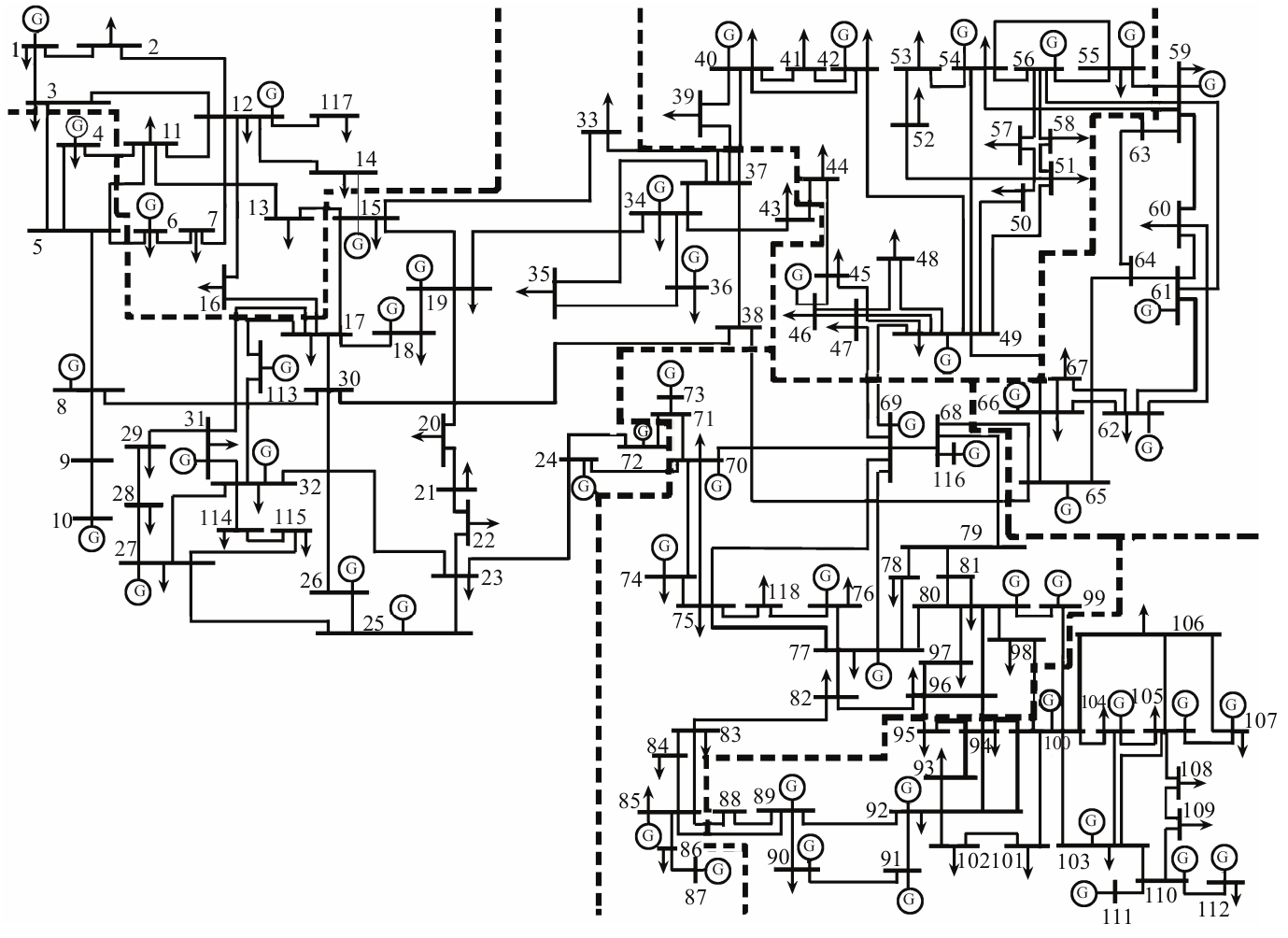}
    \setlength {\fboxsep}{1pt}
    \put(71,60) {\fcolorbox{red}{white}{\tiny \color{blue}{$52$}}} 

      \setlength {\fboxsep}{2pt}
      \put(32,70) {\fcolorbox{white}{ssBlue}{\tiny \color{black}{A1}}}
      \put(20,20) {\fcolorbox{white}{ssOrange}{\tiny \color{black}{A2}}}
      \put(41,68) {\fcolorbox{white}{ssOrange}{\tiny \color{black}{A2}}}
      \put(02,50) {\fcolorbox{white}{ssOrange}{\tiny \color{black}{A2}}}
      \put(50,70) {\fcolorbox{white}{sOrange}{\tiny \color{black}{A3}}}
      \put(92,34) {\fcolorbox{white}{Orange}{\tiny \color{black}{A4}}}
      \put(52,20) {\fcolorbox{white}{sBlue}{\tiny \color{black}{A5}}}
      \put(92,27) {\fcolorbox{white}{Blue}{\tiny \color{black}{A6}}}
      \put(75,02) {\fcolorbox{white}{Blue}{\tiny \color{black}{A6}}}

\end{overpic}
\caption{Partitioning network for the IEEE 118-bus system. There are six partitions, i.e. A1, A2, A3, A4, A5, and A6.}
\label{fig:IEEE118network}
\end{figure}
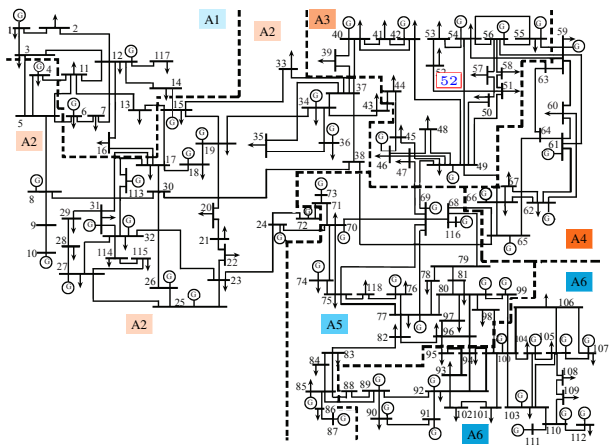

%
%

\small{}
\bibliographystyle{IEEEtran}
\bibliography{helx}

\begin{thebibliography}{10}
\providecommand{\url}[1]{#1}
\csname url@samestyle\endcsname
\providecommand{\newblock}{\relax}
\providecommand{\bibinfo}[2]{#2}
\providecommand{\BIBentrySTDinterwordspacing}{\spaceskip=0pt\relax}
\providecommand{\BIBentryALTinterwordstretchfactor}{4}
\providecommand{\BIBentryALTinterwordspacing}{\spaceskip=\fontdimen2\font plus
\BIBentryALTinterwordstretchfactor\fontdimen3\font minus
  \fontdimen4\font\relax}
\providecommand{\BIBforeignlanguage}[2]{{%
\expandafter\ifx\csname l@#1\endcsname\relax
\typeout{** WARNING: IEEEtran.bst: No hyphenation pattern has been}%
\typeout{** loaded for the language `#1'. Using the pattern for}%
\typeout{** the default language instead.}%
\else
\language=\csname l@#1\endcsname
\fi
#2}}
\providecommand{\BIBdecl}{\relax}
\BIBdecl

\bibitem{doe2003grid}
U.~DOE, ``Grid 2030: A national vision for electricity's second 100 years,''
  \emph{US DOE Report}, 2003.

\bibitem{grijalva2011prosumer}
S.~Grijalva and M.~U. Tariq, ``Prosumer-based smart grid architecture enables a
  flat, sustainable electricity industry,'' in \emph{Innovative Smart Grid
  Technologies (ISGT), 2011 IEEE PES}.\hskip 1em plus 0.5em minus 0.4em\relax
  IEEE, 2011, pp. 1--6.

\bibitem{us2004final}
U.-C. P. S. O.~T. Force, S.~Abraham, H.~Dhaliwal, R.~J. Efford, L.~J. Keen,
  A.~McLellan, J.~Manley, K.~Vollman, N.~J. Diaz, T.~Ridge \emph{et~al.},
  \emph{Final report on the August 14, 2003 blackout in the United states and
  Canada: causes and recommendations}.\hskip 1em plus 0.5em minus 0.4em\relax
  US-Canada Power System Outage Task Force, 2004.

\bibitem{wong1999ieee}
P.~Wong, P.~Albrecht, R.~Allan, R.~Billinton, Q.~Chen, C.~Fong, S.~Haddad,
  W.~Li, R.~Mukerji, D.~Patton \emph{et~al.}, ``The \text{IEEE} reliability
  test system-1996. a report prepared by the reliability test system task force
  of the application of probability methods subcommittee,'' \emph{Power
  Systems, IEEE Transactions on}, vol.~14, no.~3, pp. 1010--1020, 1999.

\bibitem{phadke2008wide}
A.~Phadke and R.~M. de~Moraes, ``The wide world of wide-area measurement,''
  \emph{Power and Energy Magazine, IEEE}, vol.~6, no.~5, pp. 52--65, 2008.

\bibitem{terzija2011wide}
V.~Terzija, G.~Valverde, D.~Cai, P.~Regulski, V.~Madani, J.~Fitch, S.~Skok,
  M.~M. Begovic, and A.~Phadke, ``Wide-area monitoring, protection, and control
  of future electric power networks,'' \emph{Proceedings of the IEEE}, vol.~99,
  no.~1, pp. 80--93, 2011.

\bibitem{xie2014dimensionality}
L.~Xie, Y.~Chen, and P.~Kumar, ``Dimensionality reduction of synchrophasor data
  for early event detection: Linearized analysis,'' \emph{Power Systems, IEEE
  Transactions on}, vol.~29, no.~6, pp. 2784--2794, 2014.

\bibitem{lim2016svd}
J.~M. Lim and C.~L. DeMarco, ``Svd-based voltage stability assessment from
  phasor measurement unit data,'' \emph{IEEE Transactions on Power Systems},
  vol.~PP, no.~99, pp. 1--9, 2015.

\bibitem{giri2012situation}
J.~Giri, M.~Parashar, J.~Trehern, and V.~Madani, ``The situation room: Control
  center analytics for enhanced situational awareness,'' \emph{Power and Energy
  Magazine, IEEE}, vol.~10, no.~5, pp. 24--39, 2012.

\bibitem{xu2013power}
W.~Xu and J.~Yong, ``Power disturbance data analytics--new application of power
  quality monitoring data,'' \emph{Proceedings of the CSEE}, vol.~33, no.~19,
  pp. 93--101, July 2013.

\bibitem{simpson2016voltage}
J.~W. Simpson-Porco, F.~D{\"o}rfler, and F.~Bullo, ``Voltage collapse in
  complex power grids,'' \emph{Nature communications}, vol.~7, 2016.

\bibitem{he2015arch}
\BIBentryALTinterwordspacing
X.~He, Q.~Ai, R.~C. Qiu, W.~Huang, L.~Piao, and H.~Liu, ``A big data
  architecture design for smart grids based on random matrix theory,''
  \emph{ArXiv e-prints}, Jan. 2015, accepted by IEEE Trans on Smart Grid.
  [Online]. Available: \url{http://arxiv.org/pdf/1501.07329.pdf}
\BIBentrySTDinterwordspacing

\bibitem{he2015corr}
\BIBentryALTinterwordspacing
X.~Xu, X.~He, Q.~Ai, and C.~Qiu, ``A correlation analysis method for power
  systems based on random matrix theory,'' \emph{ArXiv e-prints}, Jun. 2015,
  accepted by IEEE Trans on Smart Grid. [Online]. Available:
  \url{http://arxiv.org/pdf/1506.04854.pdf}
\BIBentrySTDinterwordspacing

\bibitem{he2015les}
\BIBentryALTinterwordspacing
X.~He, R.~C. Qiu, Q.~Ai, L.~Chu, and X.~Xu, ``Linear eigenvalue statistics: An
  indicator ensemble design for situation awareness of power systems,''
  \emph{ArXiv e-prints}, Dec. 2015. [Online]. Available:
  \url{http://arxiv.org/pdf/1512.07082.pdf}
\BIBentrySTDinterwordspacing

\bibitem{panteli2013assessing}
M.~Panteli, P.~Crossley, D.~S. Kirschen, D.~J. Sobajic \emph{et~al.},
  ``Assessing the impact of insufficient situation awareness on power system
  operation,'' \emph{Power Systems, IEEE Transactions on}, vol.~28, no.~3, pp.
  2967--2977, 2013.

\bibitem{endsley2011designing}
M.~R. Endsley, \emph{Designing for situation awareness: An approach to
  user-centered design}.\hskip 1em plus 0.5em minus 0.4em\relax CRC Press,
  2011.

\bibitem{Brice1982PBS}
C.~Brice and S.~K. Jones, \emph{Physically Based Stochastic Models of Power
  System Loads}.\hskip 1em plus 0.5em minus 0.4em\relax College Station, TX,
  USA: Texas A\&M Research Foundation, 1982.

\bibitem{brockwell2006introduction}
P.~J. Brockwell and R.~A. Davis, \emph{Introduction to time series and
  forecasting}.\hskip 1em plus 0.5em minus 0.4em\relax Springer Science \&
  Business Media, 2006.

\bibitem{perninge2011modeling}
M.~Perninge, V.~Knazkins, M.~Amelin, and L.~S{\"o}der, ``Modeling the electric
  power consumption in a multi-area system,'' \emph{European transactions on
  electrical power}, vol.~21, no.~1, pp. 413--423, 2011.

\bibitem{sankavaram2015incremental}
C.~Sankavaram, A.~Kodali, K.~R. Pattipati, and S.~Singh, ``Incremental
  classifiers for data-driven fault diagnosis applied to automotive systems,''
  \emph{Access, IEEE}, vol.~3, pp. 407--419, 2015.

\bibitem{yao2015large}
J.~Yao, Z.~Bai, and S.~Zheng, \emph{Large Sample Covariance Matrices and
  High-Dimensional Data Analysis}.\hskip 1em plus 0.5em minus 0.4em\relax
  Cambridge University Press, 2015.

\bibitem{Bai2010BookRMT}
Z.~Bai and J.~Silverstein, \emph{{Spectral Analysis of Large Dimensional Random
  Matrices}}, second edition~ed.\hskip 1em plus 0.5em minus 0.4em\relax
  Springer Verlag, 2010.

\bibitem{lytova2009clrforles}
A.~Lytova, L.~Pastur \emph{et~al.}, ``Central limit theorem for linear
  eigenvalue statistics of random matrices with independent entries,''
  \emph{The Annals of Probability}, vol.~37, no.~5, pp. 1778--1840, 2009.

\bibitem{najim2013gaussian}
J.~Najim, ``Gaussian fluctuations for linear spectral statistics of large
  random covariance matrices,'' \emph{arXiv preprint arXiv:1309.3728}, 2013.

\bibitem{Qiu2016BigDataLRM}
R.~C. Qiu, \emph{Big Data for Complex Network}.\hskip 1em plus 0.5em minus
  0.4em\relax CRC, 2016, ch. Large Random Matrices and Big Data Analytics.

\bibitem{guionnet2011single}
A.~Guionnet, M.~Krishnapur, and O.~Zeitouni, ``The single ring theorem,''
  \emph{Annals of Mathematics}, vol. 174, no.~2, pp. 1189--1217, 2011.

\bibitem{tao2013random}
T.~Tao and V.~Vu, ``Random matrices: Sharp concentration of eigenvalues,''
  \emph{Random Matrices: Theory and Applications}, vol.~2, no.~03, p. 1350007,
  2013.

\bibitem{marvcenko1967distribution}
V.~A. Mar{\v{c}}enko and L.~A. Pastur, ``Distribution of eigenvalues for some
  sets of random matrices,'' \emph{Sbornik: Mathematics}, vol.~1, no.~4, pp.
  457--483, 1967.

\bibitem{vanprobability}
R.~van Handel, ``Probability in high dimension,'' Princeton University, ORF 570
  Lecture Notes, June 2014.

\bibitem{qiu2013bookcogsen}
R.~Qiu and M.~Wicks, \emph{Cognitive Networked Sensing and Big Data}.\hskip 1em
  plus 0.5em minus 0.4em\relax Springer, 2013.

\bibitem{ipsen2014weak}
J.~R. Ipsen and M.~Kieburg, ``Weak commutation relations and eigenvalue
  statistics for products of rectangular random matrices,'' \emph{Physical
  Review E}, vol.~89, no.~3, 2014, \text{Art. ID 032106}.

\bibitem{qiu2015smart}
R.~Qiu and P.~Antonik, \emph{Smart Grid and Big Data}.\hskip 1em plus 0.5em
  minus 0.4em\relax John Wiley and Sons, 2015.

\bibitem{shcherbina2011central}
\BIBentryALTinterwordspacing
M.~Shcherbina, ``Central limit theorem for linear eigenvalue statistics of the
  wigner and sample covariance random matrices,'' \emph{ArXiv e-prints}, Jan.
  2011. [Online]. Available: \url{http://arxiv.org/pdf/1101.3249.pdf}
\BIBentrySTDinterwordspacing

\bibitem{johansson1998fluctuations}
K.~JOHANSSON, ``On fluctuations of eigenvalues of random hermitian matrices,''
  \emph{Duke mathematical journal}, vol.~91, no.~1, pp. 151--204, 1998.

\bibitem{guionnet2002large}
A.~Guionnet, ``Large deviations upper bounds and central limit theorems for
  non-commutative functionals of gaussian large random matrices,'' in
  \emph{Annales de l'IHP Probabilit{\'e}s et statistiques}, vol.~38, no.~3,
  2002, pp. 341--384.

\bibitem{bai2004clt}
Z.~D. Bai, J.~W. Silverstein \emph{et~al.}, ``\text{CLT} for linear spectral
  statistics of large-dimensional sample covariance matrices,'' \emph{Annals of
  Probability}, vol.~32, no.~1, p. 553, 2004.

\bibitem{anderson2006clt}
G.~W. Anderson and O.~Zeitouni, ``A \text{CLT} for a band matrix model,''
  \emph{Probability Theory and Related Fields}, vol. 134, no.~2, pp. 283--338,
  2006.

\bibitem{pan2011universality}
\BIBentryALTinterwordspacing
G.~Pan, Q.~Shao, and W.~Zhou, ``Universality of sample covariance matrices:
  \text{CLT} of the smoothed empirical spectral distribution,'' \emph{ArXiv
  e-prints}, Nov. 2011. [Online]. Available:
  \url{http://arxiv.org/pdf/1111.5420.pdf}
\BIBentrySTDinterwordspacing

\bibitem{cao2008quantum}
T.~Capo, \emph{History of quantum physics}.\hskip 1em plus 0.5em minus
  0.4em\relax Liaoning Education Press, 2008.

\bibitem{nutter1995improving}
F.~W. Nutter~Jr and P.~M. Schultz, ``Improving the accuracy and precision of
  disease assessments: selection of methods and use of computer-aided training
  programs,'' \emph{Canadian Journal of Plant Pathology}, vol.~17, no.~2, pp.
  174--184, 1995.

\bibitem{feldman2013turning}
D.~Feldman, M.~Schmidt, and C.~Sohler, ``Turning big data into tiny data:
  Constant-size coresets for k-means, pca and projective clustering,'' in
  \emph{Proceedings of the Twenty-Fourth Annual ACM-SIAM Symposium on Discrete
  Algorithms}.\hskip 1em plus 0.5em minus 0.4em\relax SIAM, 2013, pp.
  1434--1453.

\bibitem{weber2000voltage}
J.~D. Weber and T.~J. Overbye, ``Voltage contours for power system
  visualization,'' \emph{Power Systems, IEEE Transactions on}, vol.~15, no.~1,
  pp. 404--409, 2000.

\end{thebibliography}

\normalsize{}
\end{document}